\documentclass[12pt,onecolumn,draftclsnofoot,journal,letterpaper]{bare_jrnl}
\usepackage{cite}

\usepackage{graphicx}

\ifCLASSOPTIONcompsoc
\usepackage[caption=false, font=normalsize, labelfont=sf, textfont=sf]{subfig}

\else
\usepackage[caption=false, font=footnotesize]{subfig}
\fi

\usepackage{amssymb}
\usepackage{amsmath}
\newtheorem{myDef}{Definition}
\newtheorem{mytheorem}{Theorem}
\newtheorem{myproof}{Proof}

\usepackage[noend]{algpseudocode}
\usepackage{algorithmicx,algorithm}
\usepackage{array}
\usepackage{url}
\usepackage{multirow}
\usepackage{latexsym}
\usepackage{color}

\begin{document}

\title{Privacy-preserving DDoS Attacks Detection \\
Using Cross-Domain Traffic in \\
Software Defined Networks}
\label{sec:title}

\author{Liehuang Zhu,~\IEEEmembership{Member,~IEEE,}
        Xiangyun Tang,
       Meng Shen,~\IEEEmembership{Member,~IEEE,}\\
        Xiaojiang Du,~\IEEEmembership{Senior Member,~IEEE,}
        and Mohsen Guizani,~\IEEEmembership{Fellow,~IEEE}
\IEEEcompsocitemizethanks{
\IEEEcompsocthanksitem L. Zhu, X. Tang, and M. Shen are with the School of Computer Science, Beijing Institute of Technology, Beijing, China. Email: \{liehuangz, 2120161046, shenmeng\}@bit.edu.cn. Dr. Meng Shen is the corresponding author.
\IEEEcompsocthanksitem X. Du is with the Department of Computer and Information Sciences, Temple University, Philadelphia, USA. Email: dxj@ieee.org.
\IEEEcompsocthanksitem M. Guizani is with the Department of Electrical and Computer Engineering, University of Idaho, Moscow, Idaho, USA. Email: mguizani@ieee.org.
}
}

\maketitle

\begin{abstract}
\label{sec:abstract}
Existing Distributed Denial-of-Service (DDoS) attacks detections in software defined networks (SDNs)
typically only carry out detection in a single domain. In reality, abnormal traffic usually affects multiple network domains.
Thus, the cross-domain attacks detection has been proposed to improve the detection performance.
However when participating in the detection each SDNs domain needs to provide a large number of real traffic data where private information may be leaked out.
Existing multiparty privacy protection schemes often achieve privacy guarantees by sacrificing accuracy or increasing the time cost.
It is a challenging task to have both high accuracy and reasonable time consumption.

In this paper, we propose \texttt{Predis}, a privacy-preserving cross-domain attacks detection scheme for SDNs. \texttt{Predis} combines perturbation encryption and data encryption to protect privacy, and uses a computationally simple and efficient algorithm k-Nearest Neighbor (kNN) as its detection algorithm.
We also improve the kNN to achieve better efficiency.
Through theoretical analysis and extensive simulations,
we demonstrate that \texttt{Predis} is capable to achieve efficient and accurate attacks detection, while keeping sensitive information of each domain secure.
\end{abstract}


\IEEEpeerreviewmaketitle

\section{Introduction}\label{sec:Introduction}
\IEEEPARstart{S}oftware-Defined Networks (SDNs) have emerged as a new networking paradigm,
which is liberated from the vertical integration in traditional networks and gives the program and the
network their flexibility through a centralized logical network controller \cite{3}.
SDNs consists of the data plane, the control plane and the application plane.
The control plane contains some controllers that run the control logic strategy and maintains the entire network view as a logic-centric.
The controllers abstract the whole network view into network services and provide the easy-to-use interface for operators, researchers or third parties to facilitate these personnel to customize the privatization applications and realize the logical management of the network.
Users of the SDNs don't need to worry about the technical details of the underlying device, just a simple programming can realize the rapid deployment of new applications.

SDNs simplify the network management and adapt better to the current situation in which the network size continues to expand rapidly.

However the features of the centralized control and programming make SDNs susceptible to the well-known Distributed Denial-of-Service (DDoS) attacks.
For instance, the controller, which plays a crucial role in determining the functionality of each component in SDNs, is a main target of DDoS attacks \cite{3}.
A compromised controller would result in paralyzation or misbehaving of all switches under its control.

Denial of service (DoS) attack can run out of the resources of a system on the target computer, stop services and leave its normal users inaccessible.
When hackers use two or more compromised computers on the network as "puppet machine" to launch DoS attacks on a specific target, it is referred to as DDoS attacks.
The puppet machine's IP address and the structure and form of attack packet is random, making it difficult to trace the the attacker.
DDoS attacks have become a severe threat to today's Internet and the attacks make online services unavailable by overwhelming victims with traffic from multiple attackers.
With the number of businesses migrating their operations online growing dramatically, DDoS attacks can lead to significant financial losses. A recent report reveals that DDoS attacks account for 22\% of the 2015 data center downtime \cite{53}.

DDoS attacks essentially operate in three steps \cite{4}, i.e., scanning, intrusion, and attack launching. Abnormal traffic of DDoS attacks usually affects multiple paths and network domains (e.g., SDNs domains).
For ease of illustration, we analyze the stages of typical DDoS attacks using the datasets collected by the Lincoln Laboratory of MIT \cite{37}, LLDOS 1.0, as shown in Table \ref{table:Analysis of Dataset LLDOS}. Prior to the launch of the attack, abnormal traffic can be observed at the stages of scanning and intrusion.
If the victim and the puppet machines under DDoS attacks were located in different network domains,
a detection attempt restricted within a single domain would be unable to identify the attacks at their primary stages.
Thus, the involvement of multiple domains in attacks detection will help to achieve more accurate and timely detections \cite{26,38}.

\begin{table}[!t]
\renewcommand{\arraystretch}{0.9}
\caption{Analysis of LLDOS Dataset}
\label{table:Analysis of Dataset LLDOS}

\centering
\small

\begin{tabular}{c|c|c|c|c|c|c|c|}
\cline{2-8}
& \multicolumn{7}{c|}{The Number of Flows} \\
\hline
\multicolumn{1}{|c|}{Stage}&172.16.112.*&172.16.113.*&172.16.114.*&172.16.115.*&Victim&Attacker&Merge\\
 \hline
\multicolumn{1}{|c|}{Scanning}&424&328&32&296&0&1081&1081\\
\hline
\multicolumn{1}{|c|}{Intrusion}&128&0&97&0&0&332&335\\
\hline
\multicolumn{1}{|c|}{Attacking}&2&0&0&285&107465&199&107667\\
\hline
\end{tabular}
\end{table}

In the SDNs environment, the collaborative detection across multiple domains requires detailed traffic data of each domain involved, such as the content of the flow table in the latest seconds.
However this may cause serious privacy concerns on the SDNs operators, as the traffic data reveals sensitive information, such as source IP addresses, destination IP addresses, and traffic statistics \cite{40}, which have potential utility in mining network topology and network connection behaviors \cite{74,75,76}; Accordingly, SDNs operators are reluctant to share their detailed intra-domain traffic data with each other. Therefore, trade-offs between attacks detection efficiency and privacy protection should be carefully balanced in SDNs.

Many schemes of DDoS attacks detection in traditional networks have been proposed and have showed satisfying results \cite{61,77,78,79,80,81,83}.
Extensive studies on DDoS attacks in SDNs have also been done and many traffic classifiers with excellent performance have been proposed (e.g., graphic model \cite{23}, the entropy variation of the destination IP address \cite{24}, Support Vector Machine (SVM) \cite{25}).
While these DDoS attacks detection schemes are usually restricted to a single domain \cite{10}, very few studies have considered cross-domain attacks detection.
Bian et al. \cite{32} proposed a scheme for cross-domain DDoS attacks detection in SDNs using Self-organizing Map (SOM) as the traffic classifier.
The calculation in the training and the test phase is very complicated when SOM as the classifier which requires multiple vector multiplications, or more complex divition.
Secure Multi-party Computation (SMC) \cite{84,85} may enable secure cross-domain anomaly detection (e.g., secure addition protocol, secure multiplication protocol and secure compare protocol).
However, these protocols require a large amount of interaction from the participants and calculations on ciphertext, which undoubtedly consumes many of the controller's bandwidth.

Cross-domain attacks detection will lead to privacy leakage, whereas the introduction of privacy protection usually comes at a cost of excessive time consumption and low detection rate.
We should address these challenges when detecting DDoS attacks in cross-domain.
The first challenge is how to conduct cross-domain DDoS attacks detection in SDNs without revealing privacy of each network domain.
Anomaly detection classifiers require detailed traffic data, and SDNs domains do not trust each other. It is necessary that we work out the privacy issue when multiple SDNs domains work together to perform anomaly detection.
The second challenge is to ensure efficient and accurate DDoS attacks detection while well-preserving privacy.
Strong privacy protection in multi-party cooperation is often at the cost of accuracy and high time-consumption,
and it is hard to give priority to one or the other.
In the face of these dilemmas, we resort to decoupling the detection into two steps, disturbance and detection and, introducing two servers that work in collaboration to complete the detection process.

We combine digital cryptography with perturbation encryption to address the first challenge.
Transport Layer Security (TLS) is used to protect the privacy in data transmission process between two servers and SDNs controllers, and
perturbation encryption to protect that privacy is not compromised when calculating on the two servers.
The ciphertext produced by perturbation encryption, since the special design of \texttt{Predis}, can be compute directly in servers without the need to use complex security computing protocols.


With respect to the second challenge, we use the features of k-Nearest Neighbor (kNN) of calculate simple, decoupling the kNN algorithm into two steps and embedding the encryption steps into it.
After giving the training data, kNN can classify the test samples by choosing a distance measurement formula without a training phase (Euclidean distance\footnote{The formula of European distance of n-dimensional vector is $d=\sqrt {\sum ^{n}_{i=1}\left( x_{i_{1}}-x_{2i}\right) ^{2}}$.} is selected in \texttt{Predis}).
The kNN's features of easy to implement is very useful for us when we want to embed some special operations into the classifier to protect the privacy of test data.
On the other hand, kNN, as a classification algorithm, has excellent performance in accuracy, and kNN is not sensitive to outliers which can maintain high accuracy when there are some noises in dataset.
kNN as the classifier with high accuracy has been widely used in many different areas \cite{87,89,90}.

Our contributions are summarized as follows:

\begin{enumerate}
  \item We propose \texttt{Predis}, a privacy-preserving cross-domain DDoS attacks detection scheme for SDNs,
      which considers both the cross-domain DDoS attacks detection and the privacy protection in multi-party cooperation.
      \texttt{Predis} uses the features of SDNs and the improved kNN algorithm to detect DDoS attacks accurately within effective time, and combines the digital cryptography with perturbation encryption
      to provide the confidentiality and every participant's privacy.
  \item We prove the security of \texttt{Predis} by the asymptotic approach of computational security in modern cryptography.
      Through rigorous security analysis, we prove that the traffic data provided by each participant is indistinguishable for potential adversary.
  \item We conduct extensive experiments using multiple authoritative datasets to demonstrate the timeliness and accuracy of \texttt{Predis}.
      We show that our scheme not only can determine if the traffic is abnormal, but also can find abnormal traffic at the early stages of DDoS attacks.
      Results show that \texttt{Predis} is more accurate than existing detection schemes, meanwhile protecting participants' privacy.
\end{enumerate}

The rest of the paper is organized as follows.
We review the related work in Section \ref{sec:Related Work}
and introduce the thread model and security goals
in Section \ref{sec:System model and Security Model}.
We introduce the improved kNN algorithm in Section \ref{sec:Classification Method} and describe the design of \texttt{Predis} as well as the concrete calculation steps and encryption details in Section \ref{sec:Privacy-preserving Cross-domain attacks detection Scheme}.
We present the security analysis in Section \ref{sec:Safety analysis}
and the experimental results in Section \ref{sec:Experiment}. We conclude this paper in Section \ref{sec:conclusion}.

\section{Related Work}\label{sec:Related Work}

\subsection{Background of DDoS Attacks}\label{sec:Related Work-DDoS Background}

\begin{figure}[htb]
\centering
\includegraphics[width=10.5cm]{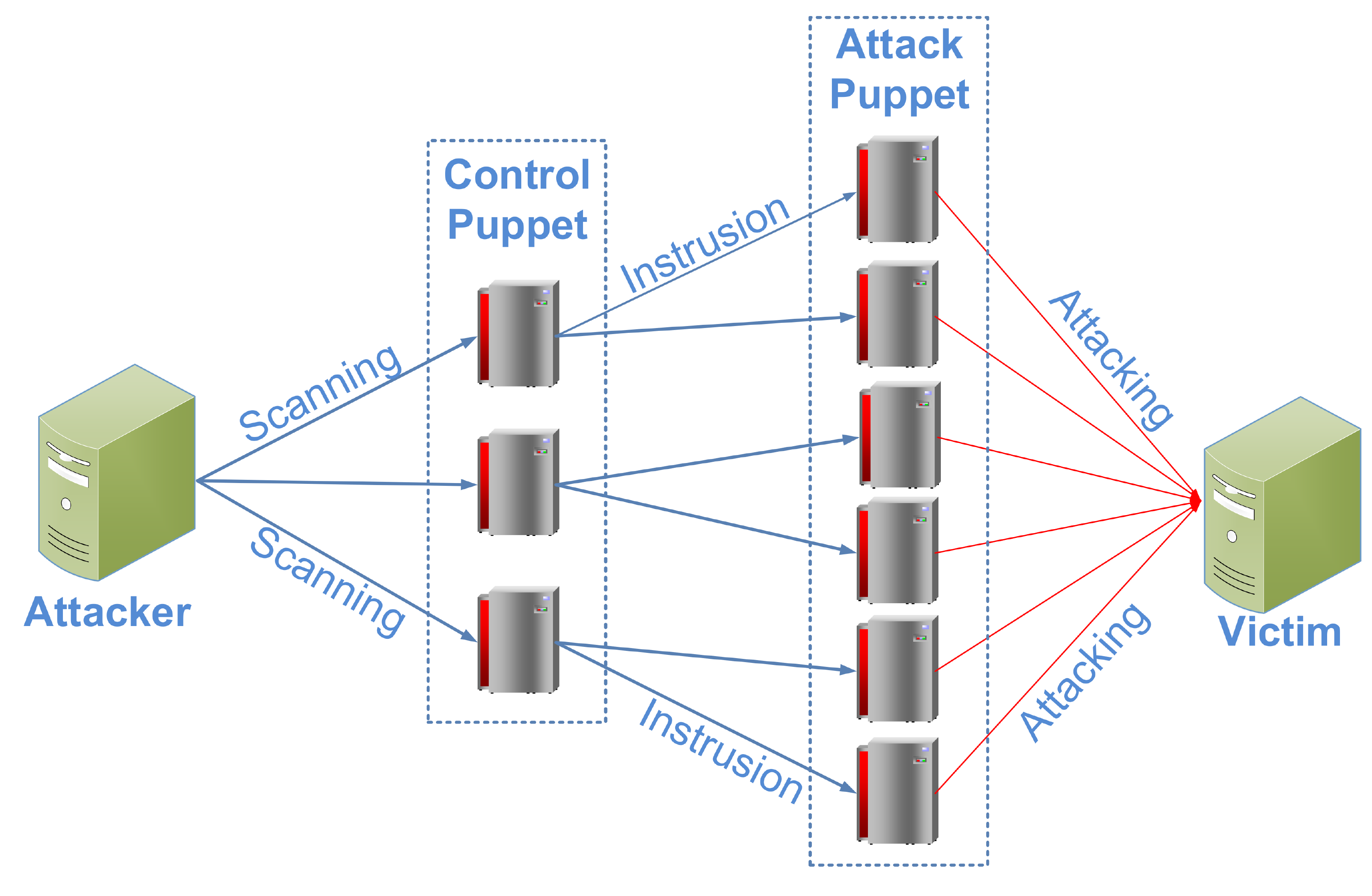}
\caption{A typical structure of DDoS Attacks}\label{fig:The schematic diagram of the DDOS}
\end{figure}

DDoS attackers can simultaneously control several computers and create an attack architecture that contains control puppets and attack puppets \cite{4}, as shown in Figure \ref{fig:The schematic diagram of the DDOS}.
The traditional attack architecture is analogous to a dumbbell-shaped structure, where an intermediate network is only responsible for data forwarding and security events and control functions are entirely done by management, while the network itself does not have the ability to detect and deal with network attacks.

\subsection{Summary of DDoS Attacks Detection Methods}\label{sec:Related Work-Summary of DDoS Attacks Detection Method in SDN}

There are numerous studies on DDoS attacks detection because of the severity and prevalence of DDoS attacks.
Here we briefly summarize the related work from two perspectives, i.e., DDoS attacks detection in conventional networks and DDoS attacks detection in SDNs, as listed in Table \ref{table:Summary of Existing DDoS attacks Detection Methods}.

\textbf{Detection in conventional networks.}
Detection approaches of DDoS attacks have been studied extensively in conventional networks, where methods such as Entropy based \cite{77}, SVM \cite{78}, Naive Bayesian \cite{80}, Neural Network \cite{81}, cluster analysis \cite{83}, Artificial Neural Network (ANN) \cite{79}, and kNN \cite{61} are used as classifiers.

\textbf{Detection in SDNs.}
The SDNs controller collects information of flow table and uses selected classifiers to classify network traffic flows as either normal or abnormal.
Based on the capability of logical centralized controller and programmability of the network, network administrators can respond to the attacks immediately.
Classic classification methods of Bayesian Networks \cite{86} and SVM \cite{25}, as well as neural networks of SOM \cite{5,32,59} and Deep Learning \cite{82} are used as traffic classifiers in SDNs.

\begin{table}[!t]
\small
\centering
\caption{Summary of Existing DDoS Attacks Detection Methods}\label{table:Summary of Existing DDoS attacks Detection Methods}

\renewcommand{\arraystretch}{0.9}
\begin{tabular}{||l|l|l||}
\hline
Network Environment&Classifiers&Reference\\
\hline
\hline
\multirow{7}{*}{Conventional Network}&Entropy based&David et al. \cite{77}\\
\cline{2-3}
&Support Vector Machine&Yusof et al. \cite{78}\\
\cline{2-3}
&Naive Bayesian&Singh et al. \cite{80}\\
\cline{2-3}
&Neural network&Hsieh et al. \cite{81}\\
\cline{2-3}
&Cluster analysis&Wei et al. \cite{83}\\
\cline{2-3}
&Artificial Neural Network&Saied et al. \cite{79}\\
\cline{2-3}
&k-Nearest Neighbor& Thwe et al. \cite{61}\\
\cline{2-3}
\hline
\hline
\multirow{7}{*}{SDNs}&\multirow{3}{*}{Self-organizing map}&Braga et al. \cite{59}\\
& &Xu et al. \cite{5}\\
& &Bian et al. \cite{32}\\
\cline{2-3}
&Support Vector Machine&Kokila et al. \cite{25}\\
\cline{2-3}
&Entropy variation of the destination IP address&Mousavi et al.\cite{24}\\
\cline{2-3}
&Deep Learning&Niyaz et al. \cite{82}\\
\cline{2-3}
& Bayesian Networks&Nanda et al. \cite{86}\\
\hline
\end{tabular}
\end{table}

The methods proposed above except Bian et al. not considered cross-domain attacks detection, naturally they didn't consider the privacy protection, and
these methods require complex calculations like vector multiplications, or more complex vector division during the testing phase (e.g., the calculation formula of Naive Bayesian is $f\left( x\right) =\prod ^{d}_{i=1}p\left( x_{i}|y\right)$\footnote{x is the test instance, d is the dimensions of test instance and y is the class mark.}).
If these methods are conducted directly for cross-domain attacks detection, secure computation protocols will be needed to solve the privacy protection.
\texttt{Predis} not only protects the privacy but also avoids the extensive interactions and calculations that are required when using secure computing protocols.

Bian et al. \cite{32} considered both cross-domain DDoS attacks detection and privacy protection.
They proposed a privacy-preserving cross-domain detection scheme, using SOM as classifier.
But their method has major complications if it came to computations, i.e., the time complexity of training a neural network and test should be $O(mn^{2})$ and $O(n^{2})$, respectively, where $n$ is the number of neurons and $m$ is the number of training samples.
The time complexity of our method in the testing phase is $O(m)$, and as a type of instance-based learning (or lazy learning), kNN don't has training phase.
In addition, they failed to consider detecting DDoS attacks at their early stages.
We attempt to find anomaly at the early stages of DDoS attacks, because if we find anomaly before the stage of attacking,
we can take countermeasures (e.g., by blocking ingress traffic with certain attack characteristics) to avoid further losses.
However, this importance has hardly been realized by most of the prior studies.
Mousavi et al. \cite{24} proposed a method to detect DDoS attacks in SDNs,
which claims that it can detect DDoS attacks within the first five-hundred packets of the attack traffic.
Given that if puppet machines and the victim were in different SDNs domains, traffic will not be reflected as being abnormal.


\subsection{Privacy-preserving in Cross-domain Detection}
A SDNs domain in \texttt{Predis} refers to a controlled domain under SDNs architecture, which is a network domain with the deployment of the SDNs techniques and can be independently controlled by operators.
The SDNs domains conduct centralized control of data forwarding.
The multiple SDNs domains described in our article collaborate and these SDNs domains may or may not be adjacent at physical or geographical location.
The control plane of SDNs domains centralized sends flow table to a specified location (i.e. computing server).
The computing server provides the DDOS detection service and return detection results controllers.
Traditional network domains for traffic forwarding is a distributed control, it can't achieve centralized control.

Privacy-preserving cross-domain attacks detection can be see as a Secure Multi-Party Computation (SMC) problem \cite{27}, which is the matter of how to safely calculate a function when no credible third party is present.
There are a hit research subject \cite{28,29,30,31} about it.

Chen et al. \cite{42} present a cryptographic protocol specially devised for privacy-preserving cross-domain routing optimization in SDNs. But these methods do not apply to the problem of cross-domain attacks detection.
Martin et al. \cite{63} investigated the practical usefulness of solutions based on SMC,
having designed optimized secure multiparty computation operations that ran efficiently on voluminous input data.
Their method may provide a new insight into the problem in this paper, but their application scenarios are not exactly the same as ours.

\texttt{Predis} uses the kNN algorithm as the classifier to perform DDoS attacks detection. The kNN technique has been employed to solve privacy-preserving problems before, and there are already several eminent secure kNN protocols.
Wong et al. \cite{71} proposed ASPE, a protocol which preserved a special type of scalar product, and constructed two secure schemes that supported kNN computation on encrypted data.
Elmehdwi et al. \cite{70} set up SkNN, which provided better security in solving the kNN query problem over encrypted database outsourced to a cloud.
Cao et al. \cite{72} proposed MRSE, which defined and solved the problem of multi-keyword ranked search over encrypted cloud data.
These secure kNN protocols focus on applying kNN to querying over encrypted data.
Secure kNN does inspire us in some ways the problem to be dealt with is DDoS attacks detection,
which demands a higher accuracy and an immediate response and is different to the querying problem over encrypted data.

Comparing with previous studies about DDoS attacks detection, \texttt{Predis} not only considers detecting DDoS attacks over multiple domains with privacy-preserving, but also attempts to detect DDoS attacks at the early stages.

\section{System Model and Threat Model}\label{sec:System model and Security Model}
In this section, we first describe the overview of the system model and the roles involved in \texttt{Predis}.
Then, we present the thread model, followed by the security goals.

\subsection{System Model}\label{sec:System model and Security Model-System Model}

\begin{figure*}
    \centering
    \subfloat[System Overview]{
       \includegraphics[width=0.48\linewidth]{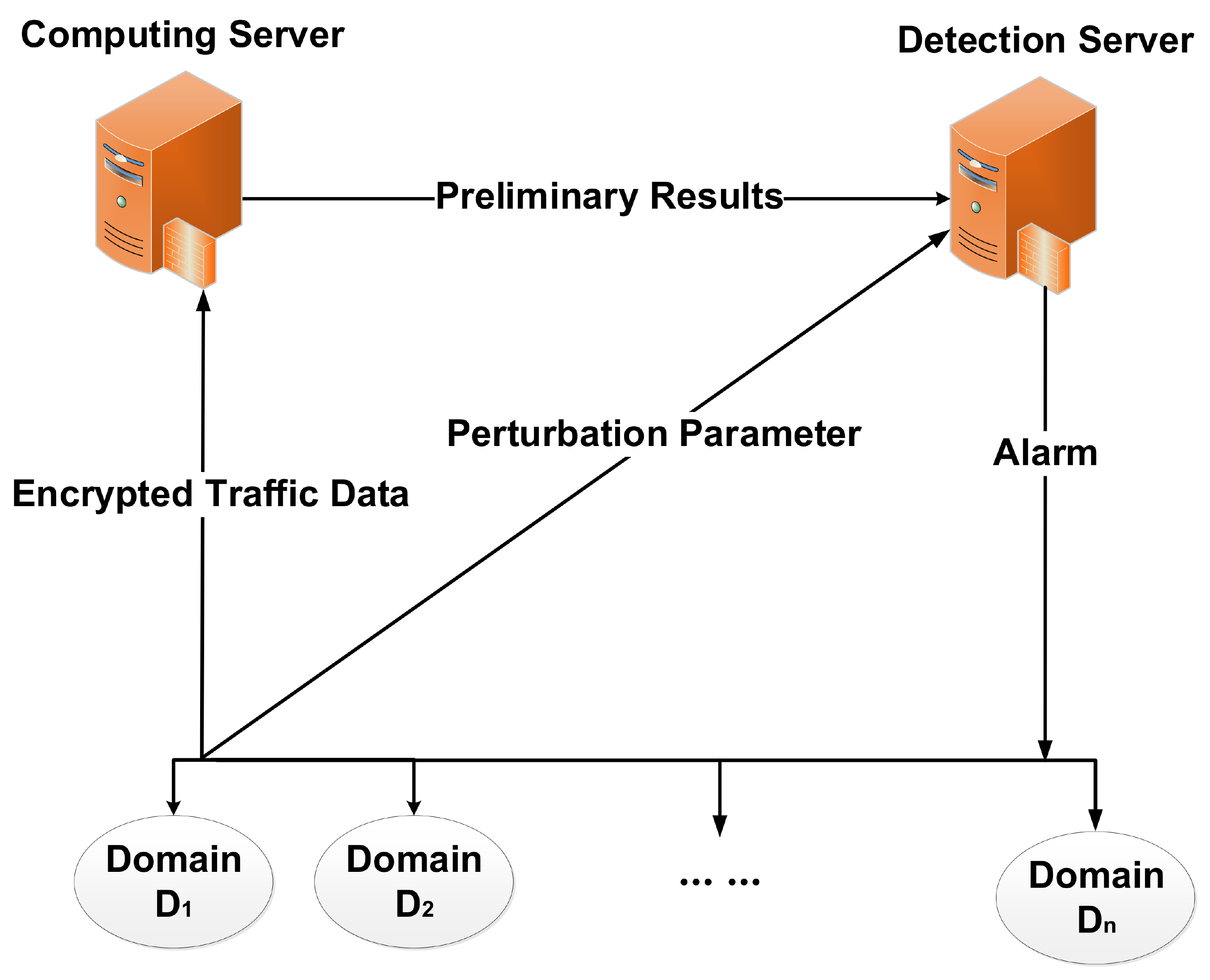}}
    \label{scheme schematic diagram-a}\hfill
    \subfloat[System Sequence Chart]{
        \includegraphics[width=0.48\linewidth]{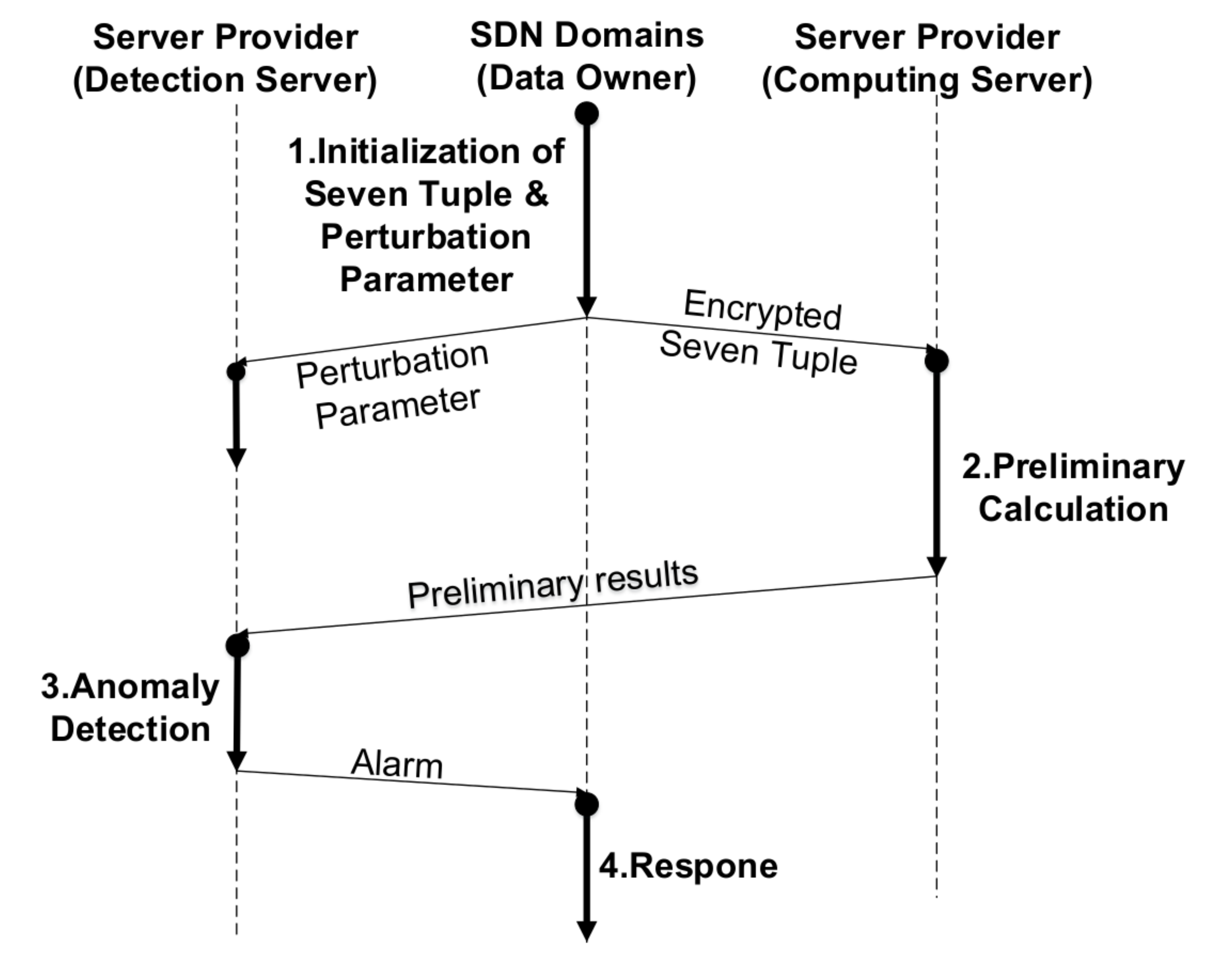}}
    \label{scheme schematic diagram-b}
    \caption{Privacy-preserving Cross-domain Attacks Detection Scheme Schematic Diagram.}
    \label{fig:scheme schematic diagram}
  \end{figure*}

\texttt{Predis} mainly contains three types of roles: Computing Server (CS), Detection Server (DS) and SDNs domains\footnote{Hereafter we refer to domains as SDNs domains unless otherwise stated.}, as exhibited in Figure \ref{fig:scheme schematic diagram}(a).
Domain $D_{n}$ is the $n$-th domain who participates in attacks detection and provides data to CS and DS, which, in turn,
provide computing and encryption services for domain $D_{n}$.

The system sequence diagram is shown in Figure \ref{fig:scheme schematic diagram}(b).
Each domain sends traffic information to CS for calculation and receives the detection results from DS.
CS provides computing service and sends the intermediate results to DS, where the latter provides detection service based on the intermediate results and replies the detection results to each domain.
Thus, CS and DS perform computation in collaboration with one another.
The details of how computing and encryption work in CS and DS will be described later in Section \ref{sec:Privacy-preserving Cross-domain attacks detection Scheme}.

\texttt{Predis} provides accurate DDoS attacks detection service for domains, where each domain is unwilling to share privacy traffic information.
Here, we give a formal definition of privacy as follows:

\begin{myDef}(Privacy). \label{def:privacy}
The information of flow table is provided by domains that participate in the detection. Specifically, privacy includes IP Source, IP Destination, Port Source, Port Destination, Length, and Flow Packets.
\end{myDef}

We define the basic operations in \texttt{Predis} of the three roles mentioned above as three functions with input and output.
Each function is designed to run on continuous inputs in real time of data partitioned into a certain time interval.
\texttt{Predis} has a set of $n$ input peers.
Input peers want to jointly compute the final result of \texttt{Predis} on their private data without the slightest relevant disclosure.
\texttt{Predis} has $m$ players called privacy peers that perform the computation of \texttt{Predis} by simulating a trusted third party (TTP) \cite{63}.
Domains are both input peers and privacy peers, while CS and DS are privacy peers.

\subsection{Threat Model}\label{sec:Security Goal and Threat Model-Threat Model}
We abstract the corss-domain privacy-preserving DDoS attacks detection problem with a threat model. In the thread model, there are two types of adversaries, namely the external adversary and the semi-honest adversary.

\textbf{\emph{External adversary}}. Adversaries that through Internet eavesdropping or data interception and other means to illegally obtain the data in the transmission process for their purposes.

\textbf{\emph{Semi-honest adversary}}. A curious participant who follows the protocol properly to fulfill service functions, but tries its best to infer sensitive or private information from intermediate results of calculation, or even colludes with other participants.

Privacy peers will set up a secure, confidential and authentic channel connecting each other to resist the external adversary.
In \texttt{Predis}, we use TLS to build this secure channel.
We adopt the semi-honest assumption for all privacy peers.
Honest privacy peers follow the protocol and do not combine their information. Semi-honest privacy peers do follow the protocol but try to infer input peers' privacy as much as possible from the values they learn, and also by combining their information.
Domains are hoping to get the correct results of attacks detection. While following the right steps, some domains may try to infer other domains' privacy for certain purpose.
CS and DS will provide the right calculation service, but may use the intermediate results generated by intermediate steps in calculation to infer and spy privacy from domains.

We assume that all privacy peers have the potential to be external adversary through eavesdropping or other methods to illegally obtain input peers' privacy. In addition to the roles included in this program, any other external eavesdropper is also the adversary we need to tackle.

\subsection{Security Goal}\label{sec:Security Goal and Threat Model-Security Goal}

The purpose of this paper is to obtain accurate cross-domain DDoS attacks detection results under the premise of privacy protection. Privacy peers may steal privacy as the external adversaries or the semi-honest adversaries.
Furthermore, privacy peers may collude with each other. In our solution, we allow one or more domains to collude with each other, with CS, and with DS.
Based on these, we make the following assumptions:
\begin{enumerate}
  \item Each domain performs function honestly but may have interest in the private information of other domains.
  \item CS or DS performs calculation process correctly but may have interest in obtaining domains' private information.
  \item CS or DS may collude with one or more domains. Semi-honest privacy peers do follow the protocol but try to infer peers' privacy as much as possible from the values they learn. Thus, CS or DS may collude with one or more domains.
  \item CS and DS do not collude with each other. In reality, DS and CS can be deployed by different operators. Operators are likely to have conflict of interests, so it is assumed that CS does not collude with DS.
\end{enumerate}

Before describing our security goals, we introduce a security definition (i.e., Definition \ref{def:security definition}), an adversarial indistinguishability experiment $PrivK_{A\cdot \pi }\left( n\right)$ as shown in Table \ref{table:indistinguishability experiment},
and a definition about negligible (i.e., Definition \ref{def:negligible definition}) for a probabilistic polynomial-time (PPT) adversary ($A$) \cite{51}.

\begin{myDef}\label{def:security definition}(Indistinguishability).
A private-key encryption scheme $\Pi = (Gen, Enc, Dec)$ has indistinguishable encryptions under an attack, if for all PPT adversaries $A$ there is a negligible function $\varepsilon$ such that
$Pr\left[ PrivK_{A\cdot \pi }\left( n\right) =1\right] =\dfrac {1} {2}+ \varepsilon (n)$, where the probability is taken over the randomness used by $A$, as well as the randomness used in the experiment.
\end{myDef}
\begin{myDef}\label{def:negligible definition}(Negligible).
A function $f$ from the natural numbers to the nonnegative real numbers is negligible if for every positive polynomial $p$ there is
an $N$ such that for all integers $n > N$ it holds that $f\left( n\right) < \dfrac {1} {p\left( n\right) }$.
\end{myDef}
\par We aim at achieving the security objective of keeping privacy of each domain.
We specify our security goals as follows:
\begin{enumerate}
  \item \emph{For CS and DS, input peers’ privacy is protected.}
  \item \emph{For domains, other input peers' privacy is protected.}
\end{enumerate}
\begin{table}[!t]
\normalsize
\centering
\small
\caption{Indistinguishability Experiment $PrivK_{A\cdot \pi }\left( n\right)$}\label{table:indistinguishability experiment}
\renewcommand{\arraystretch}{0.9}
\begin{tabular}{|m{15cm}|}
\hline
\textbf{The indistinguishability experiment $PrivK_{A\cdot \pi }\left( n\right)$:}\\
1. A key $k$ is generated by running $Gen\left(1^{n}\right)$.\\
2. The adversary $A$ is given input $1^{n}$, and outputs a pair of messages $m_{0}$, $m_{1}$ of the same length.\\
3. A uniform bit $b\in\left\{0,1\right\}$ is chosen, and then a ciphertext $c\leftarrow Enc\left(m_{b}\right)$ is computed and given to $A$.\\
4. $A$ outputs a bit $b'$.\\
5. The output of the experiment is defined to be $1$ if $b' = b$, and $0$ otherwise. In the former case, we say that $A$ succeeds.\\
\hline
\end{tabular}
\end{table}

\section{Classification Method}\label{sec:Classification Method}

To adapt to the proposed privacy protection scheme, we design a classifier by the kNN algorithm, decoupling it into two steps and embedding the encryption steps into it.
In this section, we will introduce the details of how traffic classification is carried out.

\subsection{Improved kNN as Classifier}\label{sec:Classification Method-Improved kNN as Classifier Algorithm}


In general, kNN is implemented by linear scanning \cite{33}.
In linear scanning, we need calculate every distance between the test data and training data, and than sort and find the nearest $k$ instances.
When the training dataset is very large, the computation will be very time-consuming.

KD-tree is a balanced binary tree that divides the entire attribute space into specific $d$ parts according to the number of attributes of the dataset, and then carries out relevant query operations in a specific space.
Best Bin First (BBF) is an optimization algorithm for querying on the KD-tree,
the main idea of which is to sort the nodes in the "querying path", and the retroactive checking is always performed from the best-priority tree node.
Using KD-Tree to store training dataset and searching with BBF not only don't need to calculate every distance between test data and training data, but also improve the efficiency.
Readers interested in KD-Tree or BBF can read literature \cite{91}, because these are not the focus of \texttt{Predis}, we don't detail describe it here.

In Section \ref{sec:System model and Security Model}, we have introduced the system model where CS has the training dataset, and DS provides detection service.
So, we decoupled kNN into two steps:
In the first step, CS builds a KD-tree based on the training dataset, and calculates the preliminary results of the distance between the test data and the ordered training data.
The second step, DS gets the preliminary results and finds the nearest $k$ instances by using BBF.
The time complexity of kNN with linear scan is $O\left( n\log k\right)$, and with BBF the time complexity is $O(n)$.
When the dataset is large, the time consumption shorten by BBF is very impressive.
Main steps are shown in Algorithm \ref{athm:improved kNN}.

\begin{algorithm}[!t]
\renewcommand\baselinestretch{1.2}\selectfont
\small
\caption{Improved kNN Algorithm}\label{athm:improved kNN} 
\begin{algorithmic}[1]
\Require Training datasets $D_{t}$, test instance $t$, timeout limit $L$.
\Ensure Detection result $y$.
    \State Building KD-tree based on the dimensions of training data in CS.
    \State calculating the preliminary results from the ordered training data in CS.
    \State Removing the perturbation from preliminary results to get the correct distance in DS.
    \State $t$ as the root is added into traversal queue.
    \State Initializing a queue $Q_{k}$.
\While{traversal queue is not null \textbf{and} $(L != 0)$}
	\State $node$ $\leftarrow$ traversal queue' top.
    \State Get the $distance$ between $node$ and $t$ distance.
    \If{$distance$ $<$ $Q_{k}$'s top}
        \State remove $Q_{k}$'s top.
        \State Insert $node$ to $Q_{k}$.
    \EndIf
    \If{$node$'s n-th dimension's value $\leqq$ $t$'s n-th dimension's value}
        \State $node$'s left children enters traversal queue.
        \State traverse right subtree.
    \Else
        \State $node$'s right children enters traversal queue.
        \State traverse left subtree.
    \EndIf
\EndWhile
    \State Getting the detection result $y$ by queue $Q_{k}$.
\State \Return $y$.
\end{algorithmic}
\end{algorithm}


\subsection{Feature Selection}\label{sec:Classification Method:Feature Selection}

The proposed DDoS attacks detection scheme is based on the flow table obtained from SDNs controllers.
As there are a lot of redundant information inside, which affects not only the detection efficiency but also the results, we extract feature information from the flow table.
Normal traffic is generally interactive because the purpose of it is to obtain or provide services,
but the number of ports and source IP addresses will increase significantly when DDoS attacks occur.
One of the other features of a DDoS attacks is source IP spoofing, which usually results in a lot of traffic with a small number of packets with a small number of bytes.
Normal flow usually has many packets, and the number of flow's bytes is larger.
So we calculate the median of packets per flow and bytes per flow to reinforce this feature instead of the mean, because the mean is possible to smooth this feature.

To quantify these characteristics, five parameters are used in the feature selection module,
including MPF, MBF, PCF, GOP and GSI, which are elaborated as follows:
\begin{enumerate}
  \item
  Median of Packets per Flow (MPF), which describes the number of packets' median in every $n$ flows.
  We rank the flows $X=\left\{ X_{1},X_{2},\ldots ,X_{n}\right\}$ in ascending order based on the number of \emph{packets} per flow, and then compute the median value according to Formula \eqref{equ:median}.
  \begin{equation}\label{equ:median}
  M\left( X\right) =\begin{cases} X_{\left( n+1\right) / 2},\ \qquad  \qquad n\ is\ odd;\\
   \dfrac {X_{\left (n / 2\right )}+X_{\left( n/2 +1\right)}} {2},\ otherwise.\end{cases}
  \end{equation}
  \item
  Median of Bytes per Flow (MBF), which describes the number of bytes' median in every $n$ flows.
  We rank the flows in ascending order based on the number of \emph{bytes} per flow, and then compute the median value.
  \item
  Percentage of Correlative Flow (PCF), which describes the number of flows with interactive features in every $n$ flows.
  We define flow $X$ as $X = (srcIP = A, dstIP = B)$ and flow $Y$ as $Y = (srcIP = B, dstIP = A)$, where $X$ and $Y$ use the same protocol. $PCF=inactN/n$, where $inactN$ is the number of $X$ in addition to the number of $Y$.
  \item
  Growth of Ports (GOP), which describes the growth rate of the number of ports within a fixed time.
  $GOP=portN /t$, where $t$ is the fixed time interval and $portN$ is the number of port growth.
  \item
  Growth of Source IP Addresses (GSI), which describes the growth rate of the number of source IP addresses within one fixed time.
  $GSI=srcIPN/t$, where $srcIPN$ is the number of source IP addresses growth.
\end{enumerate}

\section{Privacy-Preserving Cross-domain Attacks Detection Scheme}\label{sec:Privacy-preserving Cross-domain attacks detection Scheme}

In this section, we describe the workflow of \texttt{Predis} and detail the processes of how to combine privacy protection in DDoS attacks detection.

\subsection{Encryption in Data Transmission Process}\label{sec:scheme-Digital Envelope Encryption}


To avoid traffic data being leaked in transmission process, we leverage the TLS \cite{73}.

TLS is a security protocol that provide secure connections between two applications to communicate across a network to exchange data and is designed to prevent eavesdropping and tampering.
Before the application layer protocol communication, TLS protocol has completed the encryption algorithm, the communication key agreement and server authentication.
Application layer protocol can be created transparently on the TLS protocol.
TLS consists of three basic steps:
The client asks and verifies the public key to the server;
Both parties negotiate to generate session key;
Both parties use the session key for encrypted communications.
TLS for information transfer across a network is considered safe and reliable up to now.

\begin{algorithm}[!t]
\renewcommand\baselinestretch{1.2}\selectfont
\small
\caption{Traffic Pretreatment in Each SDNs Domain}\label{athm:traffic pretreatment} 
\begin{algorithmic}[1]
\Require Traffic data of flow table.
\Ensure Ciphertext passed to DS $C_{DS}$, ciphertext passed to CS $C_{CS}$.
\State Initialize seven tuple set $T$ by feature selection module's formula and traffic data of flow table.
\While{1}
    \For{$\forall t\in T$}
        \For{$i\leftarrow0 < 6$}
            \State Random generation disturbance parameter $r=\left\{ 0,1\right\} ^{231}$.
            \State $\Delta t_{i}=t_{i}+r$.
        \EndFor
    \EndFor
    \For{$\forall \Delta t\in \Delta T$}
        \State $C_{CS}\leftarrow Enc_{skc}\left( \Delta t\right) $.
    \EndFor
    \For{$\forall $r$\in R$}
        \State $C_{DS}\leftarrow Enc_{skd}\left( r\right) $.
    \EndFor
    \State \Return $C_{DS}$ to DS.
    \State  \Return $C_{CS}$ to CS.
\EndWhile
\end{algorithmic}
\end{algorithm}
\subsection{Traffic Pretreatment in Domains}\label{sec:scheme-Traffic Pretreatment in domains}

In traffic pretreatment, domains need collect traffic and abstract each piece of traffic information as a seven tuple, followed by generating perturbation parameter respectively.
The obtained seven tuple is encrypted by the perturbation parameter. Ultimately, domains transmit the encrypted seven tuple to CS, the perturbation parameter to DS, as shown in Algorithm \ref{athm:traffic pretreatment}.

Domains collect and transmit traffic information every 3 seconds, since an overly long interval would cause the network paralyzed before the attacks are detected,
while an overly short one would make the resource utilization of detection module too high to handle other requests in the controller, which can cause heavy load on the link between the controller and its switches.
\begin{figure*}[t]
    \centering
    \subfloat[SDNs Flow Table Content]{
       \includegraphics[width=0.45\linewidth]{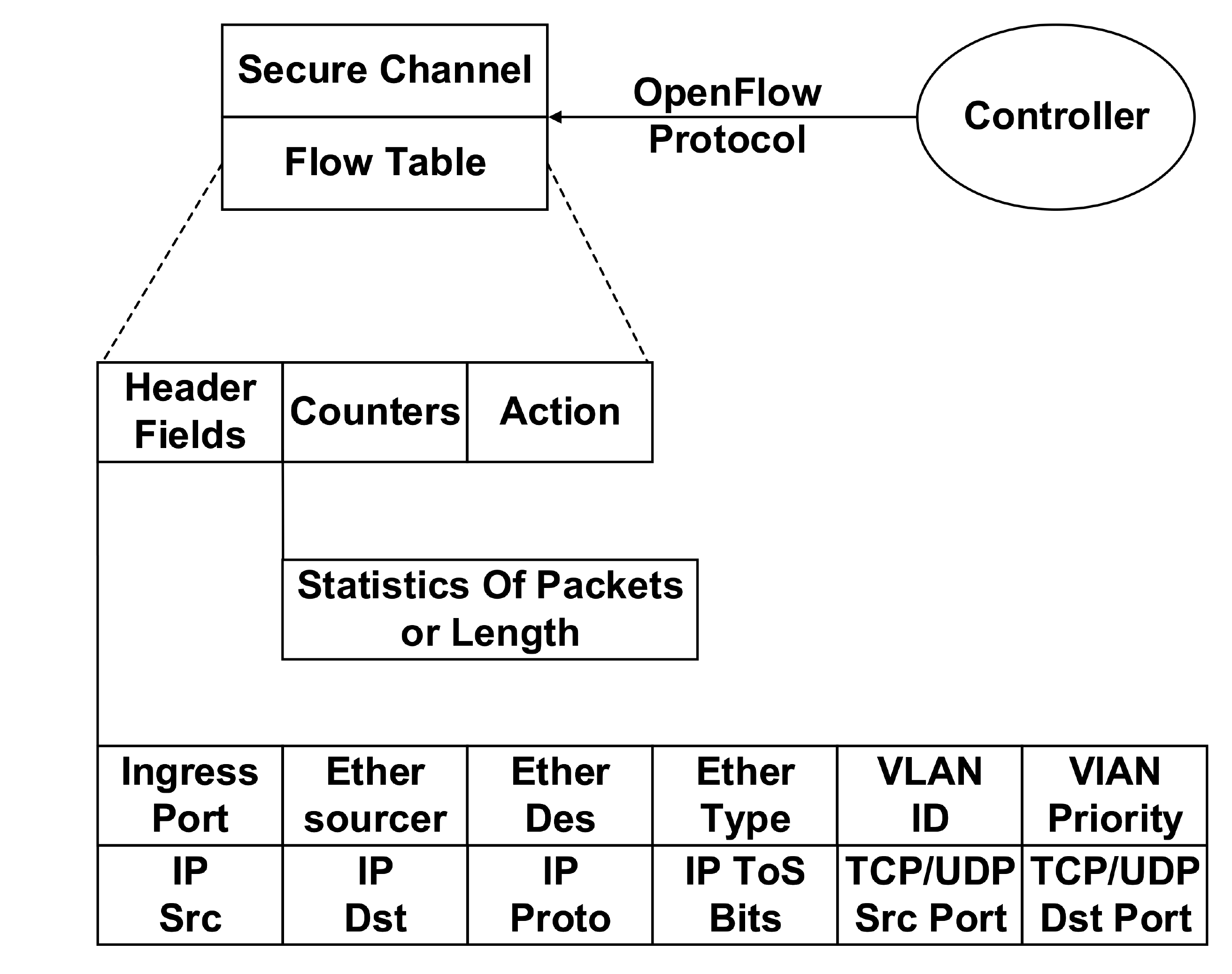}}
       \hfill
    \subfloat[Traffic Pretreatment]{
        \includegraphics[width=0.45\linewidth]{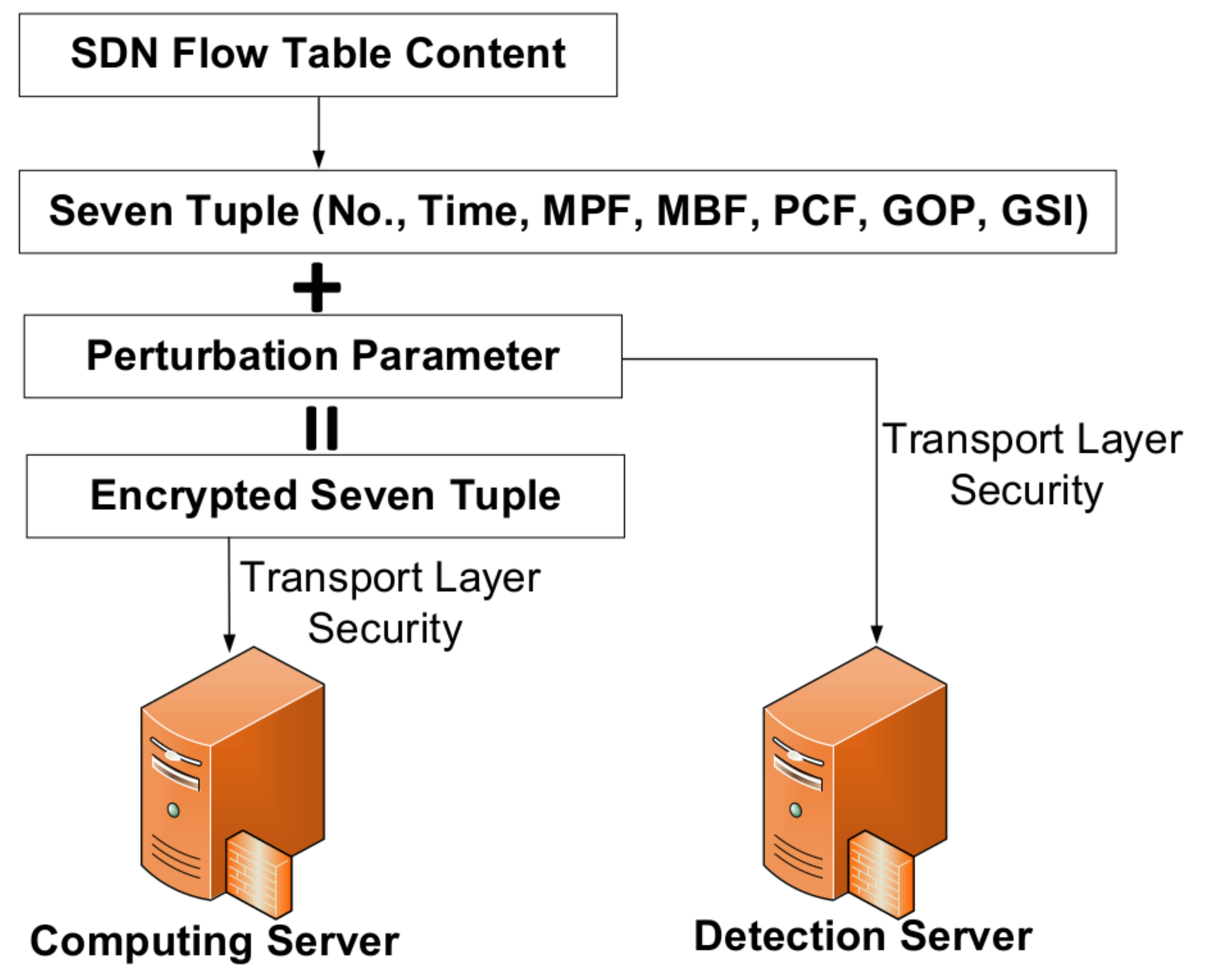}}
    \caption{The Format of SDNs Flow Table and Traffic Pretreatment.}
    \label{fig:Traffic Pretreatment in domains}
\end{figure*}

As mentioned in Section \ref{sec:Classification Method:Feature Selection}, the information needed in detection are source IP, destination IP, source port, destination port, flow bytes, and flow packets.
Domains go through the process written in the flow table with the equations described in Section \ref{sec:Classification Method:Feature Selection} and calculate the MPF, MBF, PCF, GOP and GSI.
We define a seven tuple as $\langle$Serial Number, Time, MPF, MBF, PCF, GOP and GSI $\rangle$.
The functions of Serial Number and Time are similar to the primary key in \emph{relational database}. It is a label that uniquely identifies this flow table item generated by domains. In experiments we set the Serial Number as number $n$ for $n$-th domain participating in detection, and Time as the timestamp of the flow table item.
Each attribute in the seven tuple is stored as a binary of 33 bits (we add an additional bit as an overflow flag),
and the total length of the seven tuple is 231 bits.
Attributes less than 33 bits will be filled with 0 in front. 

In each domain, the disturbance parameter is added to the seven tuple. Using the TLS, domains securely deliver the encrypted seven tuple to CS, and, the disturbance parameter to DS.
Flow table content is shown in Figure \ref{fig:Traffic Pretreatment in domains}(a) and the schematic diagram of the traffic pretreatment in domains is shown in Figure \ref{fig:Traffic Pretreatment in domains}(b).

\subsection{Preliminary Calculation in CS}\label{sec:scheme-Preliminary Calculation in CS}
Upon receiving the encrypted seven tuple, CS calculates preliminary seven tuple used for attacks detection.
Then, CS sends the results to DS by using TLS.

\begin{figure}[t]
    \centering
    \subfloat[Preliminary Calculation in CS]{
       \includegraphics[width=0.50\linewidth]{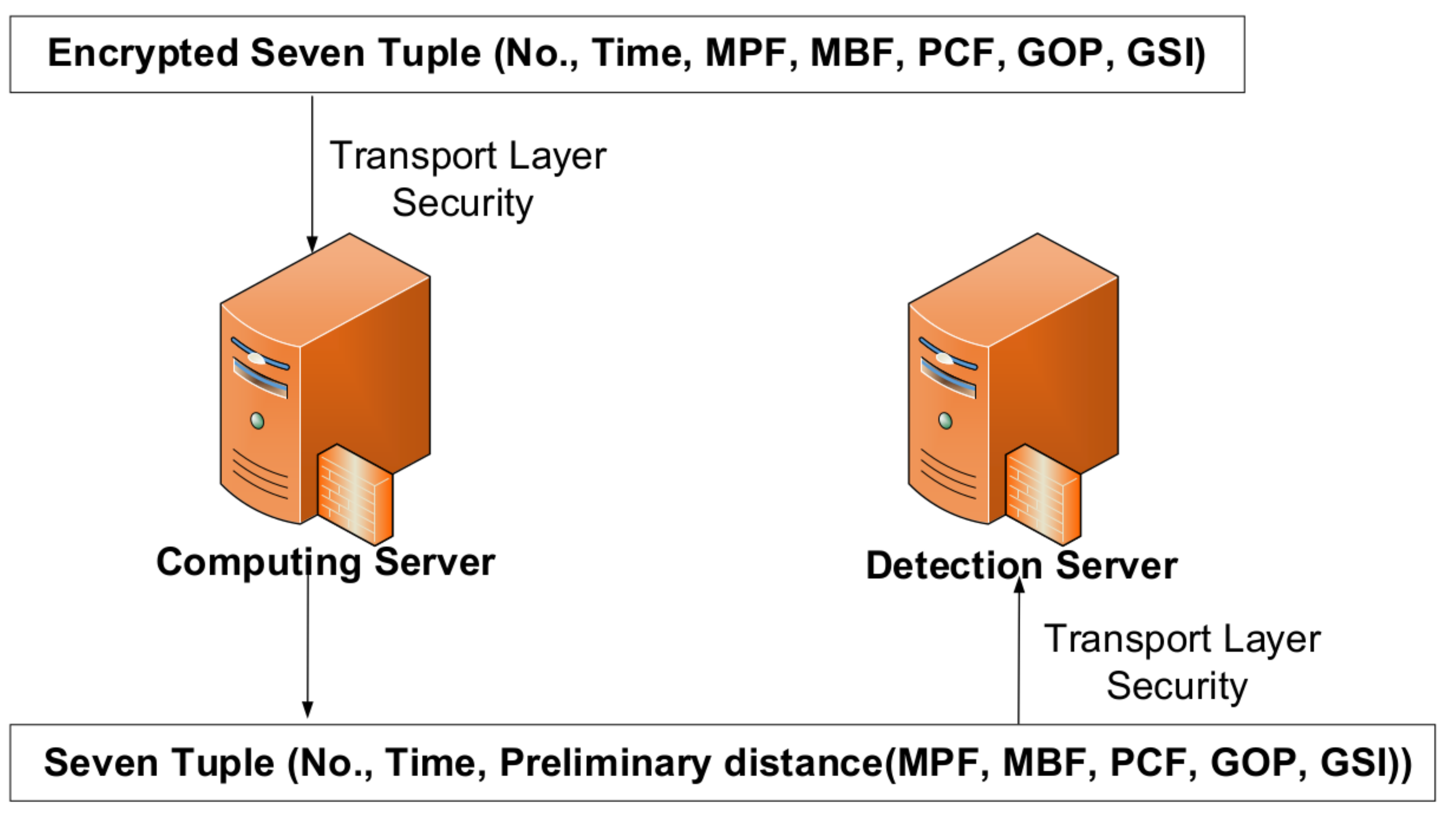}}
       \hfill
    \subfloat[Attacks Detection in DS]{
        \includegraphics[width=0.45\linewidth]{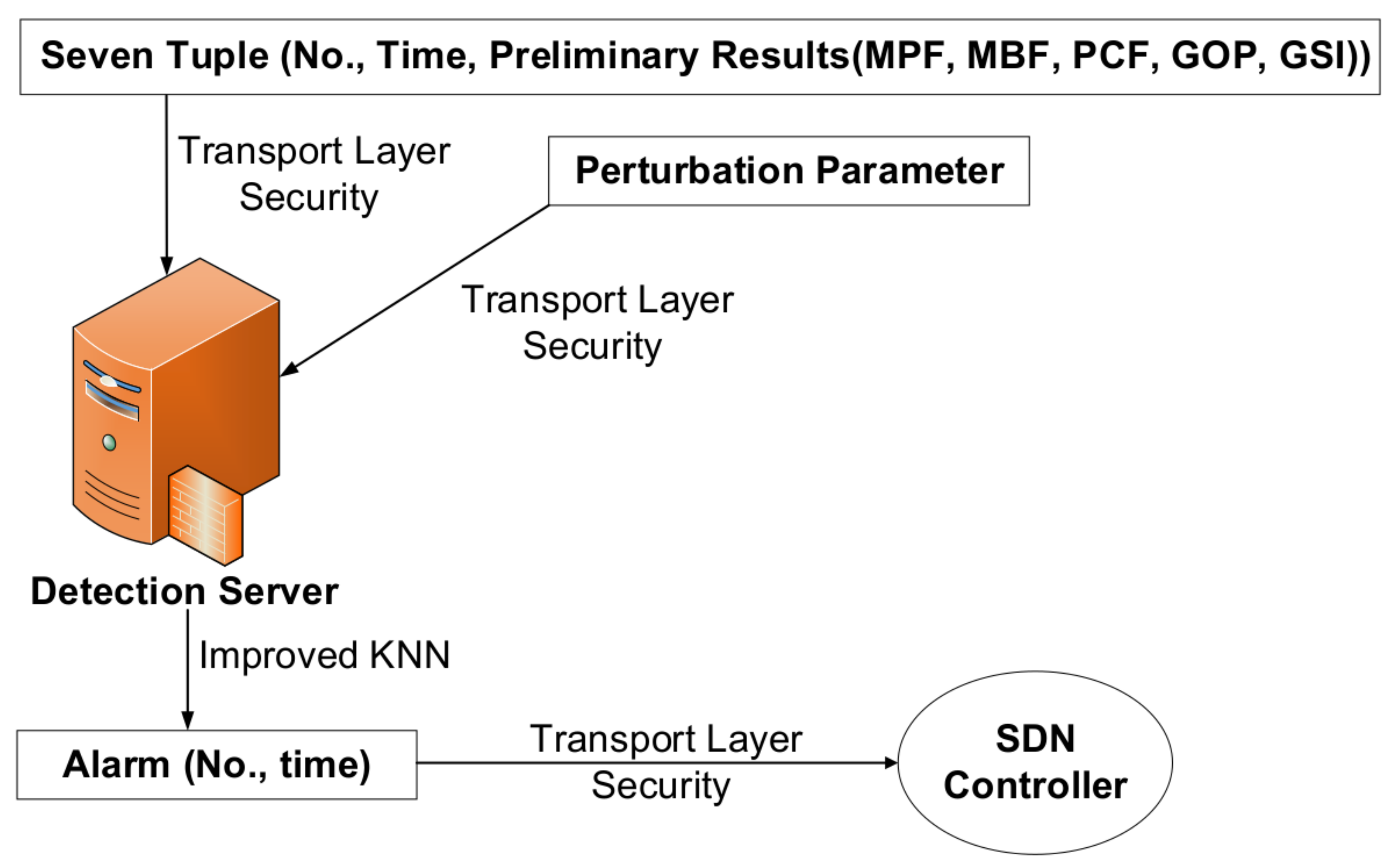}}
    \caption{Schematic Diagram of Preliminary Calculation and Attacks Detection.}
    \label{fig:DS and CS's schematic}
\end{figure}

The calculation process in CS is exhibited in Algorithm \ref{athm:preliminary calculation}.
\texttt{Predis} employs the kNN for attacks detection, computing distance is thus an important step.
We calculate the preliminary results of the distance between the test data and the training data.
CS calculates preliminary results in the received encrypted seven tuple directly and obtains the distance between the disturbance data and the training data.
We leave the work of removal of the perturbation and get the exact result of distance to DS.
The schematic diagram is shown in Figure \ref{fig:DS and CS's schematic}(a).

\begin{algorithm}[!t]
\renewcommand\baselinestretch{1.2}\selectfont
\small
\caption{Preliminary Calculation in CS}\label{athm:preliminary calculation} 
\begin{algorithmic}[1]
\Require Training data $D_{t}$, set of encrypted seven tuple $\bigtriangleup T$.
\Ensure $C_{DS}$.
\While{1}
    \For{$\forall t\in T$}
        \For{$\forall d_{t}\in D_{t}$}
            \State $PreR = D_{t} - t$.
        \EndFor
    \EndFor
    \For{$\forall preR\in PreR$}
        \State $C_{DS}\leftarrow Enc_{skc}\left( preR\right) $.
    \EndFor
    \State  \Return $C_{DS}$ to DS.
\EndWhile
\end{algorithmic}
\end{algorithm}

\subsection{Attacks Detection in DS}\label{sec:scheme-attacks detection in DS}

The attacks detection process in DS is abstracted in Figure \ref{fig:DS and CS's schematic}(b) and exhibited in Algorithm \ref{athm:Anomaly Detection}.
Upon receiving domains' perturbation parameters and CS's preliminary results of the distance,
DS removes the perturbation from preliminary results to get the correct distance.
The improved kNN uses the correct distance to get correct detection results.
Finally, if the classifier finds DDoS attacks, DS will returns the alarm (Serial Number and Time) to domains. Domains' operator will respond appropriately after receiving an alarm.

The calculation results of CS are training data $D_{t}$ minus the perturbed seven tuple $\Delta T$ ($D_{t} - \Delta T$).
Since DS has the perturbation parameters $R$, it can get the correct distance for attacks detection by using the perturbation parameters with the calculation results of CS ($D_{t} - \Delta T - R$) subtracted.
Finally, the improved kNN calculates the results of DDoS attacks detection.

\begin{algorithm}[!t]
\renewcommand\baselinestretch{1.2}\selectfont
\small
\caption{Attacks Detection in DS}\label{athm:Anomaly Detection} 
\begin{algorithmic}[1]
\Require Seven tuple of preliminary calculation's result $PreR$, perturbation parameter $R$.
\Ensure Alarm from detection $A$.
\While{1}
    \For{$\forall preR\in PreR$}
        \For{$\forall r\in R$}
            \State $fanlR = preR - r$.
        \EndFor
        \State Doing attacks detection by using $FanlR$.
        \State get the detection result $A$ by Algorithm \ref{athm:improved kNN}.
    \EndFor
    \State  \Return $A$ to domains.
\EndWhile
\end{algorithmic}
\end{algorithm}

\section{Security Analysis}\label{sec:Safety analysis}

As described in Section \ref{sec:Security Goal and Threat Model-Security Goal}, our security goal is to protect the privacy of each input peers.
\texttt{Predis} uses TLS to protect privacy in data transmission process.
TLS is generally accepted as secure for data transfer across a network.
Besides the correct data receiver, any external eavesdropper cannot eavesdropping and tampering data.
Therefor, we don't analysis the security of TLS here.

A scheme is secure if any PPT adversary succeeds in breaking the scheme with at most negligible probability \cite{51}.
In other words, PPT adversary succeeds in the indistinguishable experiment which is showed in Table \ref{table:indistinguishability experiment} with at most negligible probability, we could say that the scheme is secure.
It is the Asymptotic Approach in model cryptography to prove the security of a scheme and we use this idea in the following.

In this section, we would use the idea of the Asymptotic Approach and present the proofs by showing the indistinguishability in the following two situations.
That is, for CS and DS, input peers' private information is indistinguishable, and for domains, other input peers' private information is indistinguishable.

Before formal security analysis, we first set out the meaning of each representation:
$T$ is private information mentioned in Section \ref{sec:Security Goal and Threat Model-Security Goal};
$\bigtriangleup T$ is the set of encrypted $T$;
$R$ is the disturbance parameter of each domains;
$PreR$ is preliminary calculation's result output by CS.

As for domains, what the legal data domains have is their own data and detection result.
When they try to gain privacy from others, such as when through means of external eavesdropper, privacy is indistinguishable.

As for CS, the legal data it has is
$\bigtriangleup T = T + R$.
We construct an encryption scheme $\Pi_{CS}$ as shown in Table \ref{table:construction pics}.
We give a theorem (Theorem \ref{def:CS security}) for it and prove it.
If CS does not collude with domains, it is merely a ciphertext-only eavesdropping adversary at this point.
If CS colludes with one or more domains, this attack would be a chosen-plaintext attack (CPA) in construction $\Pi_{CS}$' s encryption scheme.
In a ciphertext-only attack, the only thing the adversary needs to do is eavesdrop on the public communication channel over which encrypted messages are sent \cite{51}.
In the chosen-plaintext attack the adversary is assumed to be able to obtain encryptions and/or decryptions of plaintexts/ciphertexts of its choice \cite{51}.
Chosen-plaintext adversary has more useful information than the ciphertext-only adversary, and chosen-plaintext adversary is harder to prevent than the ciphertext-only adversary.
When we stopped the chosen-plaintext adversary, we can stopped the ciphertext-only adversary.

By \emph{Proof} \ref{pro:CS security proof}, we demonstrate Construction $\Pi_{CS}$ is a CPA-secure private-key encryption scheme for messages of length $l$.
Thus, input peers' private information is indistinguishable for CS.

\begin{mytheorem}\label{def:CS security}
If $G$ is a pseudorandom function, then Construction $\Pi_{CS}$ is a CPA-secure private-key encryption scheme for messages of length $l$.
\end{mytheorem}

\begin{myproof}\label{pro:CS security proof}
Let $\overline{\Pi} = (\overline{Gen}, \overline{Enc}, \overline{Dec})$ be an encryption scheme that is exactly the same as $\Pi_{CS}$, except that a truly random function $g$ is used in place of $G_{k}$.
Fix an arbitrary ppt adversary $A$, and let $q(n)$ be an upper bound on the number of queries that $A\left( 1^{n}\right)$ makes to its encryption oracle. We show that there is a negligible function $\varepsilon (n)$ as follow and prove this by reduction.
\begin{equation}
\left| Pr\left[ PrivK_{A\cdot \Pi_{CS} }^{cpa}\left( n\right) =1\right] - Pr\left[ PrivK_{A\cdot \overline{\Pi_{CS}} }^{cpa}\left( n\right) =1\right] \right|=\varepsilon (n)
\end{equation}

We use $A$ to construct a distinguisher $D$ for the pseudorandom function $G$. The distinguisher $D$ is given oracle access to some function $O$, and its goal is to determine whether this function is ``pseudorandom" (i.e., equal to $G_{k}$ for uniform $k\in \left\{ 0,1\right\} ^{n}$ or ``random").
To do this, $D$ emulates experiment $PrivK^{cpa}$ for $A$ in the manner described below, and observes whether $A$ succeeds or not. If $A$ succeeds then $D$ guesses that its oracle must be a pseudorandom function, whereas if $A$ does not succeed then $D$ guesses that its oracle must be a random function.

$D$ runs in polynomial time since $A$ does. The key points are as follows:

If $D$'s oracle is a pseudorandom function, then the view of $A$ when running as a subroutine by $D$ is distributed identically to the view of $A$ in experiment $PrivK_{A\cdot \Pi_{CS} }^{cpa}\left( n\right)$.
This is because, in this case, a key $k$ is chosen uniformly at random and then every encryption is carried out by choosing a uniform $r$, and setting the ciphertext equal to $\left\langle r,G(K)+m\right\rangle$, exactly as in Construction $\Pi_{CS}$.

If $D$'s oracle is a random function, then the view of $A$ when running as a subroutine by $D$ is distributed identically to the view of $A$ in experiment
$PrivK_{A\cdot \overline{\Pi_{CS}} }^{cpa}\left( n\right)$. This can be seen exactly as above, with the only difference being that a uniform function is used instead of $G_{k}$.

\begin{equation}\label{equ:security analysis}
\begin{split}
Pr \left[ C=c|M=m\right]& =\Pr\left[ M+K=c|M=m\right] =\Pr\left[ m+K=c\right] \\
&=\Pr\left[ K=c-m\right] =\dfrac {1} {2^{l}}\\
\end{split}
\end{equation}

Through Formula (\ref{equ:security analysis}), we know $Pr\left[ PrivK_{A\cdot \overline{\Pi_{CS}} }^{cpa}\left( n\right) =1\right] = \dfrac {1}{2}$. Combining the above and the assumption that $G$ is a pseudorandom function, there exists a negligible function $\varepsilon (n)$ for which $Pr\left[ PrivK_{A\cdot \Pi_{CS} }^{CS}\left( n\right) =1\right] =\dfrac {1} {2}+ \varepsilon (n)$. From the Definition \ref{def:security definition}, we have proved Construction $\Pi_{CS}$ is a CPA-secure private-key encryption scheme for messages of length $l$.$\hfill\blacksquare$

\end{myproof}

\begin{table}[!t]
\normalsize
\centering
\small
\caption{Construction of $\Pi_{CS}$}\label{table:construction pics}
\renewcommand{\arraystretch}{0.9}
\begin{tabular}{|m{15cm}|}
\hline
Let $G$ be a pseudorandom function. Define a private-key encryption scheme for messages of length $l$.\\
\hline
\textbf{\emph{Gen}}: on input $1^{n}$, choose uniform $K\in \left\{ 0,1\right\} ^{n}$ and output it.\\
\textbf{\emph{Enc}}: on input a key $K\in \left\{ 0,1\right\} ^{n}$ and a massage $m\in \left\{ 0,1\right\} ^{l(n)}$, output the ciphertext.\\
\qquad\qquad\qquad\qquad\qquad\quad$C:=\left\langle r,G(K)+m\right\rangle$\\
\textbf{\emph{Dec}}: on input a key $K\in \left\{ 0,1\right\} ^{n}$ and a ciphertext $c\in \left\{ 0,1\right\} ^{l(n)}$, output the plaintext message.\\
\qquad\qquad\qquad\qquad\qquad\quad$M:=\left\langle G(K)-s\right\rangle$\\
\hline
\end{tabular}
\end{table}

As for DS, the legal data it has is $R$ and $PreR$.
Since $PreR = D_{t} - \bigtriangleup T$ and $\bigtriangleup T = T + R$, DS has legal data $D_{t} - T$.
We construct an encryption scheme $\Pi_{DS}$ as shown in Table \ref{table:construction pids}.
If DS does not collude with one or more domains, it is a ciphertext-only eavesdropping adversary at this point for encryption scheme $\Pi_{DS}$.
If DS colludes with one or more domains: having no other CS's private key, they would be unable to get other input peers' $\bigtriangleup T$.
DS owns perturbation parameter $PreR$.
Therefore, this situation is CPA-secure in encryption scheme $\Pi_{DS}$, which is stated in Theorem \ref{def:DS security}.
By \emph{Proof} \ref{pro:DS security proof}, we demonstrate that input peers' private information is indistinguishable for DS.

\begin{table}[!t]
\normalsize
\centering
\small
\caption{Construction of $\Pi_{DS}$}\label{table:construction pids}
\renewcommand{\arraystretch}{0.9}
\begin{tabular}{|m{15cm}|}
\hline
\textbf{\emph{Gen}}: choose $K\in \left\{ 0,1\right\} ^{n}$ and output it.\\
\textbf{\emph{Enc}}: on input a key $K\in \left\{ 0,1\right\} ^{n}$ and a massage $m\in \left\{ 0,1\right\} ^{l(n)}$, output the ciphertext.\\
\qquad\qquad\qquad\qquad\qquad\quad$C:=\left\langle K-m\right\rangle$\\
\textbf{\emph{Dec}}: on input a key $K\in \left\{ 0,1\right\} ^{n}$ and a ciphertext $c\in \left\{ 0,1\right\} ^{l(n)}$, output the plaintext message.\\
\qquad\qquad\qquad\qquad\qquad\quad$M:=\left\langle K-s\right\rangle$\\
\hline
\end{tabular}
\end{table}

\begin{mytheorem}\label{def:DS security}
The encryption scheme $\Pi_{DS}$ is a CPA-secure private-key encryption scheme for messages of length $l$.
\end{mytheorem}

\begin{myproof}\label{pro:DS security proof}
Through Formula (\ref{equ:security analysis2}), we learn that $Pr\left[ PrivK_{A\cdot \Pi_{DS} }^{DS}\left( n\right) =1\right] =\dfrac {1} {2}+ D(K)$ in ciphertext-only attack, where $D(K)$ is the probability of guessing training set.
\begin{equation}\label{equ:security analysis2}
\begin{split}
Pr \left[ C=c|M=m\right]& =\Pr\left[ K-M=c|M=m\right]=\Pr\left[ K-m=c\right] \\
&=\Pr\left[ K=c+m\right]\\
\end{split}
\end{equation}

In general, there are 2500 records in training data set for kNN, each of them being 32*5=160 bits in our scheme.
Thus, $D\left( k\right) =\dfrac {1} {2^{2500\times 160}}$. With the idea of asymptotic approach, we consider $D\left( k\right)$ negligible.

In chosen-plaintext attack, $Pr\left[ PrivK_{A\cdot \Pi_{DS} }^{DS}\left( n\right) =1\right]$ could be equal with $\dfrac {1} {2}+ \dfrac {q(n)} {2^{2500\times 160}}$, where $q(n)$ is the number of queries to the Oracle.

If $Q(n)$ is a polynomial about n, $\varepsilon (n)*Q(n)$ is still negligible.
In encryption scheme $\Pi_{DS}$, $q(n)$ is a polynomial about the number of collusive domians, so $\dfrac {q(n)} {2^{2500\times 160}}$ is negligible.
Encryption scheme $\Pi_{DS}$ has indistinguishable encryptions under a chosen-plaintext attack.$\hfill\blacksquare$


\end{myproof}


\section{Evaluation}\label{sec:Experiment}
This section evaluates \texttt{Predis} in terms of accuracy, expansibility, time consumption, and compatibility.

\subsection{Preliminary}\label{sec:Experiment-Preliminary}
\textbf{Dataset}.
Since simulating attack scenarios has a major defect in terms of traffic diversity, we employ five sets of public traffic traces for our experiments, including the \emph{CAIDA ``DDoS attacks 2007" traces} \cite{46},
the {\emph{CAIDA Anonymized 2008 Internet traces} \cite{67},
the \emph{2000 DARPA LLDOS 1.0 and LLDOS 2.0.2 traces}\cite{37},
the {\emph{1999 DARPA traces} \cite{37},
and the \emph{KDD Cup 1999 traces} \cite{50}.
Besides, we deployed a DDoS attacks experiment and captured relevant traffic traces for our experiments.
The file format of these datasets is .pcap which pertains every packets' detail.
We parse these .pcap files by flow statistics
to simulate the flow table collected by a controller in SDNs.

Using the combinations of these traces,
we define three datasets for experiments.

\begin{table}[!t]
\centering
\small
\caption{Statistics of Dataset 1}\label{table:Dataset 1}

\renewcommand{\arraystretch}{0.9}
\begin{tabular}{||c|c|c|c|c|c|c||}
\hline
\multirow{2}{*}{Domains}&\multicolumn{2}{c|}{1999 DARPA}&\multicolumn{2}{c|}{LLDOS 1.0}&\multicolumn{2}{c||}{LLDOS 2.0.2}\\
\cline{2-7}
&Packets&Flows&Packets&Flows&Packets&Flows\\
\hline
\hline
172.16.112.*& 1573963&427056 & 1237&1104&376&354			\\
\hline
172.16.113.*& 585996& 236122 &	338&328&255&	254		\\
\hline
172.16.114.*&835099 & 311354&	344&258&24&	16		\\
\hline
172.16.115.*&5090 &3838&2428&1336&632&611			\\
\hline
131.84.1.31(Victim)&35645 & 17830&	108509&107465&2100&2003		\\	
\hline
202.77.162.213(Attacker)&52606 &24498 &	3722&3174&1724&1634	\\	
\hline	
Domains Merge& 2502403& 1020698&		116578&107667&5111&	560	\\
\hline
\end{tabular}
\end{table}

\textbf{Dataset 1}.
The 1999 DARPA and 2000 LLDOS traces were collected from the same network topology.
Thus, in Dataset 1, we use the 1999 DARPA traces as normal traffic, and use the 2000 LLDOS traces as anomaly traffic.
We segment domains by the IP address segment.
Victims and attackers are located in different domains.
Statistics of Dataset 1 are shown in Table \ref{table:Dataset 1}.

\textbf{Dataset 2}.
All traffic in the CAIDA were collected from both directions of an OC-192 Internet backbone link by CAIDA's equinix-chicago monitor.
Thus, in Dataset 2, we use the CAIDA Anonymized 2008 Internet traces as normal traffic, and use the CAIDA ``DDoS attacks 2007" traces as anomaly traffic.

\textbf{Dataset 3}.
We used Python and Scapy\footnote{Scapy is a python library used for interactive packet.} to achieve the simulation of synchronous (SYN) flood attack.
To simulate DDoS attacks, we used $5$ hosts to launch SYN flood attacks against a host, and then collected 5 minutes of abnormal traffic on this victim host.
To obtain the abnormal traffic as clean as possible, when collecting abnormal traffic, we closed all the applications on the victim host.
We collected another 45 minutes normal traffic in this victim host when there is no attacks.
Statistics of Dataset 3 are shown in Table \ref{table:Statistics of dataset 3}.

\begin{table}[!t]
\centering
\small
\caption{Statistics of Dataset 3}\label{table:Statistics of dataset 3}

\renewcommand{\arraystretch}{0.9}
\newcommand{\tabincell}[2]{\begin{tabular}{@{}#1@{}}#2\end{tabular}}
\begin{tabular}{c|c|c|}
\cline{2-3}
 &Packets&Flows \\
\cline{2-3}
\hline
\multicolumn{1}{|c|}{Anomaly traffic} & 114214 & 79440 \\
\hline
\multicolumn{1}{|c|}{Normal traffic} & 941904 & 18582 \\
\hline
\end{tabular}
\end{table}

In addtion, the KDD Cup 1999 traces are used alone to evaluate performance of \texttt{Predis} in part of compatibility.
Statistics of the KDD Cup 1999 traces are shown in Table \ref{table:Statistics of KDD Cup 1999}.

\begin{table}[t]
\centering
\small
\caption{Statistics of Cross Validation Using Dataset 2}\label{table:Dataset 2}

\renewcommand{\arraystretch}{0.9}
\begin{tabular}{||c|p{2cm}<{\centering}|p{2cm}<{\centering}|p{2cm}<{\centering}|p{2cm}<{\centering}||}
\hline
\multirow{2}{*}{Domains}&\multicolumn{2}{c|}{CAIDA Anonymized 2008}&\multicolumn{2}{c||}{CAIDA DDoS attacks 2007}\\
\cline{2-5}
&Packets&Flows&Packets&Flows\\
\hline
\hline
Sample 1& 1370524&243997 & 435428&	63093		\\
\hline
Sample 2& 1377329& 243724 &	426769&	62574	\\
\hline
Sample 3& 1286528&237956 &	445796&	63749		\\
\hline
Sample 4& 1299980&234976 &	431176&	63349		\\
\hline
Sample 5&1340870 &246775 &	398453&59467	\\	
\hline
Sample 6& 1338945&243518 &	413624&	61057\\	
\hline	
\end{tabular}
\end{table}


\textbf{Cross-Validation}.
To evaluate the performance difference between \texttt{Predis} and others, we employ cross-validation for each dataset.

\emph{Dataset 1}.
To protect traffic characteristics of DDoS attacks of per phase in LLDOS, we do not partition LLDOS's data.
When LLDOS 1.0 is used as the training dataset, LLDOS 2.0.2 will be used as the test dataset.
When LLDOS 2.0.2 is used as the training dataset, LLDOS 1.0 will be used as the test dataset.
DARPA1999 traces always act as background traffic, selecting 50\% of which to create the training dataset and the remaining 50\% for validation.

\emph{Dataset 2}.
As network traffic data is time-dependent, we divide traces on a time basis.
When performing cross validation, we divide both the CAIDA Anonymized 2008 Internet traces and the CAIDA ``DDoS attacks 2007" traces into 6 partitions per 3 seconds.
Each time we take one of them as the test dataset, the remainder as the training dataset.
Cross validation statistics of Dataset 2 are shown in Table \ref{table:Dataset 2}.

\begin{table}[!t]
\centering
\small
\caption{Statistics of KDD Cup 1999 Traces}\label{table:Statistics of KDD Cup 1999}

\renewcommand{\arraystretch}{0.9}
\newcommand{\tabincell}[2]{\begin{tabular}{@{}#1@{}}#2\end{tabular}}
\begin{tabular}{|c|c|c|c|c|}
\hline
Type of Attacks & Normal & DoS & Prob & U2R\\
\hline
\hline
Number of Connections & 972781 & 391502 & 4107 & 218 \\
\hline
\end{tabular}
\end{table}

\textbf{Methods to Compare}.
\texttt{Predis} is a privacy-preserving cross-domain detection in SDNs.
To evaluate the performance of \texttt{Predis} in a comprehensive way, we select three methods as the state-of-the-art for comparison, i.e., SVM, SOM, and PSOM.
Kokila et al. \cite{25} leverage SVM to perform DDoS attacks detection in SDNs whereby high accuracy rate has been achieved.
Braga et al. \cite{59} use SOM to perform DDoS attacks detection in SDNs.
PSOM is a cross-domain DDoS detection scheme using SOM as the classifier and introduces privacy-preserving proposed by Bian et al. \cite{32}.
Besides, liner kNN (kNN) is implemented to clear the improvement towards kNN in \texttt{Predis} in term of speed.
We disable the privacy-preserving function in \texttt{Predis} and name it as PkNN, and Naive Bayes (NB) is also implemented to have a better view.


\textbf{Comparison Criteria}.
The fundamental goal of attacks detection is accuracy (i.e., identifying more anomalies in the ground truth and avoiding false alarms) \cite{69}. We use precision ($\frac{\#correctly\ divided\ into\ attack\ flows}{\#correctly\ and\ falsely\ divided\ into\ attack\ flows}$) and recall ($\frac{\#correctly\ divided\ into\ attack\ flows}{\#all\ attack\ flows}$) to measure the detection accuracy.

\subsection{Evaluation of Classifier Performance }

\emph{(1) Selection of The Best $k$ Value}.

To find the best $k$ value for the improved kNN in \texttt{Predis},
we observe the changes in time consumption, precision, and recall, when $k$ increases from 5 to 35 gradually,
where the privacy-preserving component in \texttt{Predis} is temporarily disabled.
Subsequently, we determine and select the best $k$ value for our scheme.
The experimental data for evaluation is the Sample 1 in Dataset 2.
The training dataset size is 2400, including 1200 normal traffic instances and 1200 abnormal traffic instances,
whose scale remains unchanged in the following experiments.

Experimental results are depicted in Figure \ref{fig:best K value}, where the vertical coordinates are changes in the $k$ value. Figure \ref{fig:best K value}(a) and (b) exhibit the evaluation results of precision (recall) and time consumption, respectively.
We can find that an appropriate $k$ value lies between 20 and 25, where \texttt{Predis} achieves relatively high accuracy without introducing heavy time overhead.
Thus, we choose the $k$ value as 23 in the following experiments.

\begin{figure*}
\begin{minipage}[t]{0.48\textwidth}
    \centering
    \includegraphics[height=5cm]{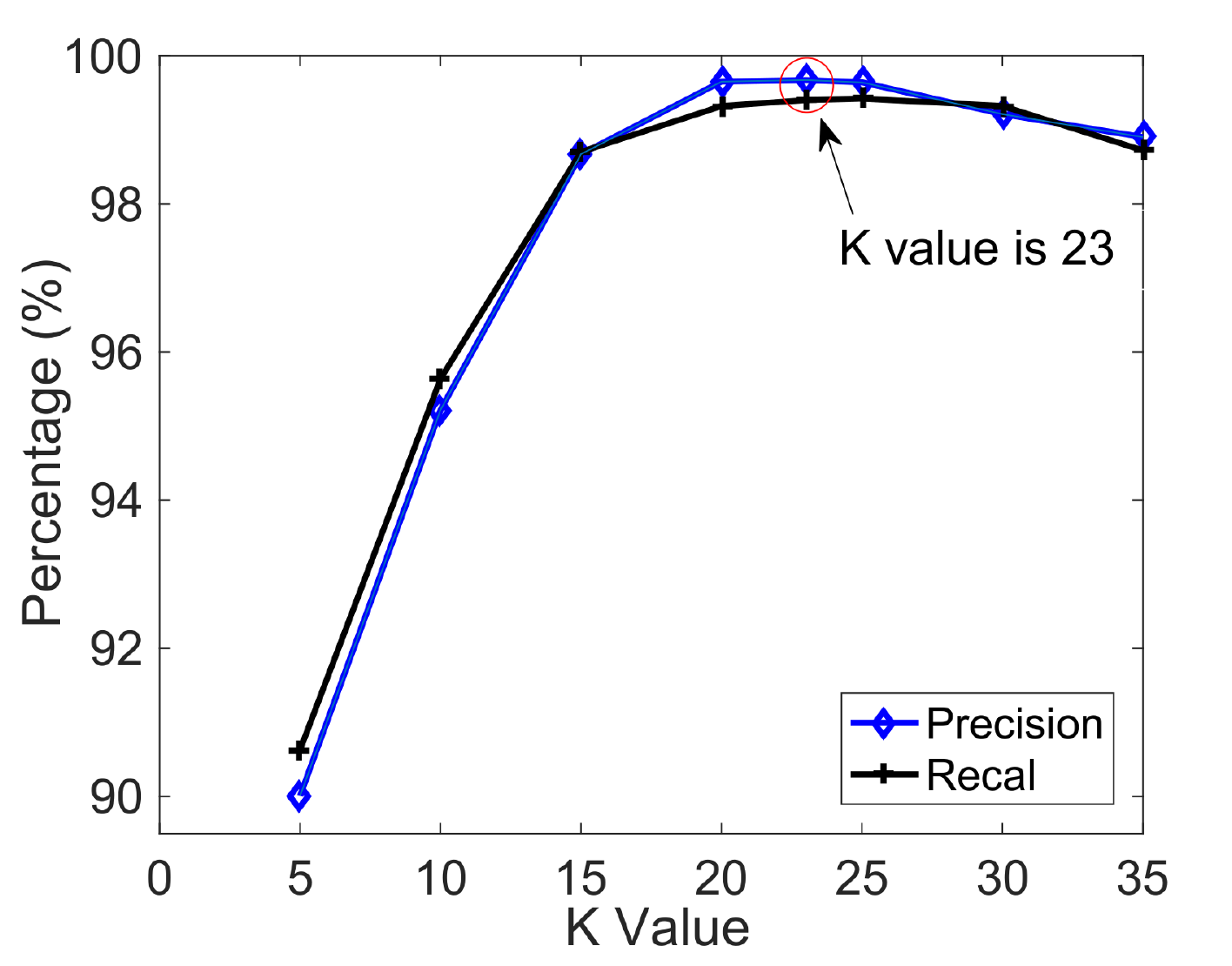}\\
    \scriptsize{(a) Precision}
\end{minipage}%
\quad
\begin{minipage}[t]{0.48\textwidth}
    \centering
    \includegraphics[height=5cm]{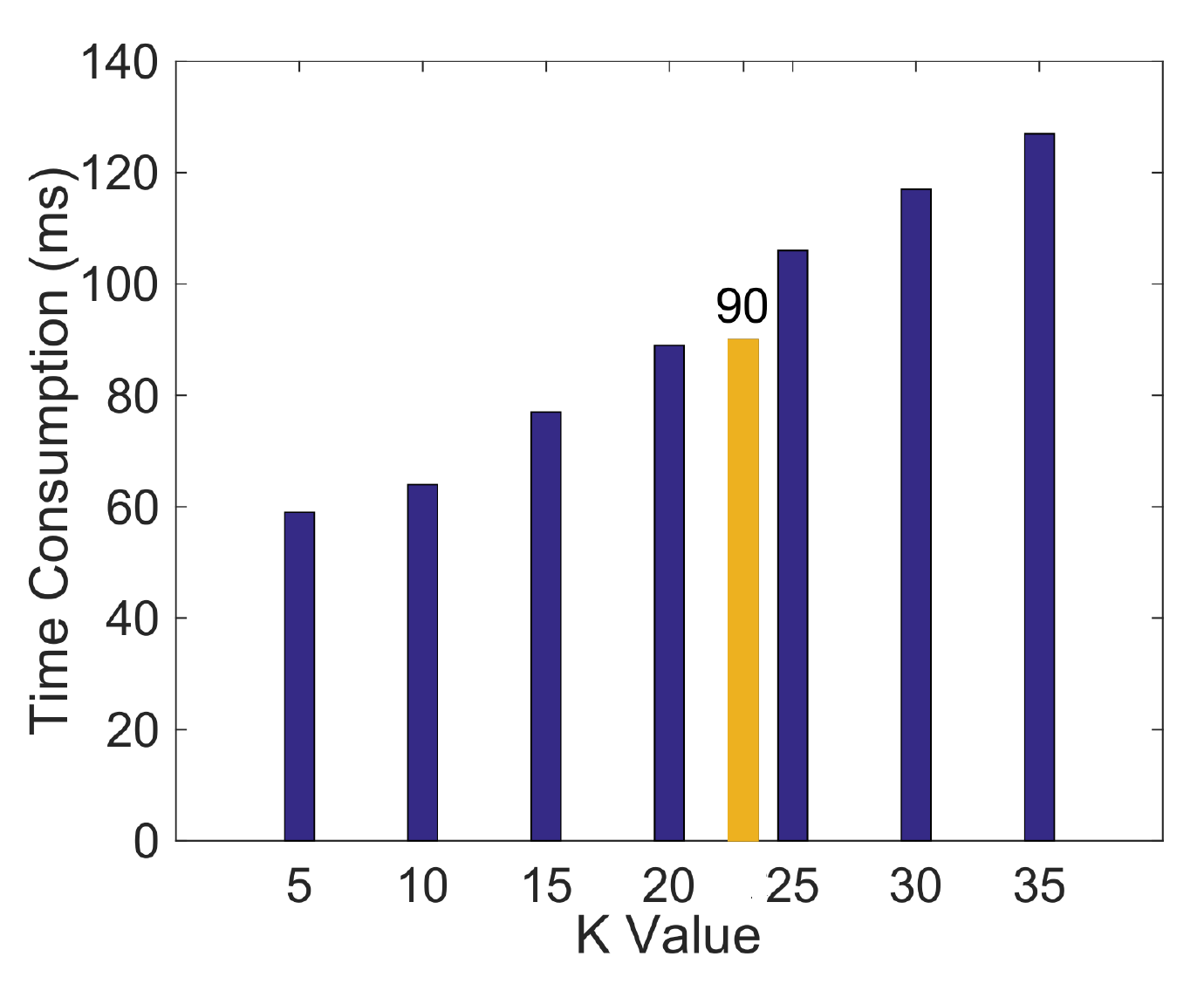}\\
    \scriptsize{(b) Time Consumption}
\end{minipage}%
 \caption{Precision and Time Consumption by Varying $k$.}
    \label{fig:best K value}
\end{figure*}

\emph{(2) Classifier Performance}.

To better comprehend the performance of the classifier in \texttt{Predis} in speed and accuracy, we conduct a comparison between \texttt{Predis}, PkNN, kNN, PSOM, SOM, SVM and NB by using Dataset 3, which results is depicted in Figure \ref{fig:Comprehensive Evaluation of the Classifier in Predis}.
It can be seen from the Figure \ref{fig:Comprehensive Evaluation of the Classifier in Predis}(a), compared to other algorithms, kNN has a higher accuracy, but its speed has no advantage.

Although we use some algorithms (KD-tree and BBF) to improve its speed, it is still not the least time consumption algorithm as shown in the Figure \ref{fig:Comprehensive Evaluation of the Classifier in Predis}(b) by the PkNN speed.
An important reason for choosing kNN as the classifier in \texttt{Predis} is that kNN is easy to calculate (cf., Section \ref{sec:Classification Method-Improved kNN as Classifier Algorithm}) which facilitates embedding the encryption steps into it, and another reason is that kNN has relatively higher accuracy which have been confirmed in this experiment.
Besides, \texttt{Predis} and PSOM have relatively high time consumption because of the privacy protection process.

\begin{figure*}
\begin{minipage}[t]{0.48\textwidth}
    \centering
    \includegraphics[height=5cm]{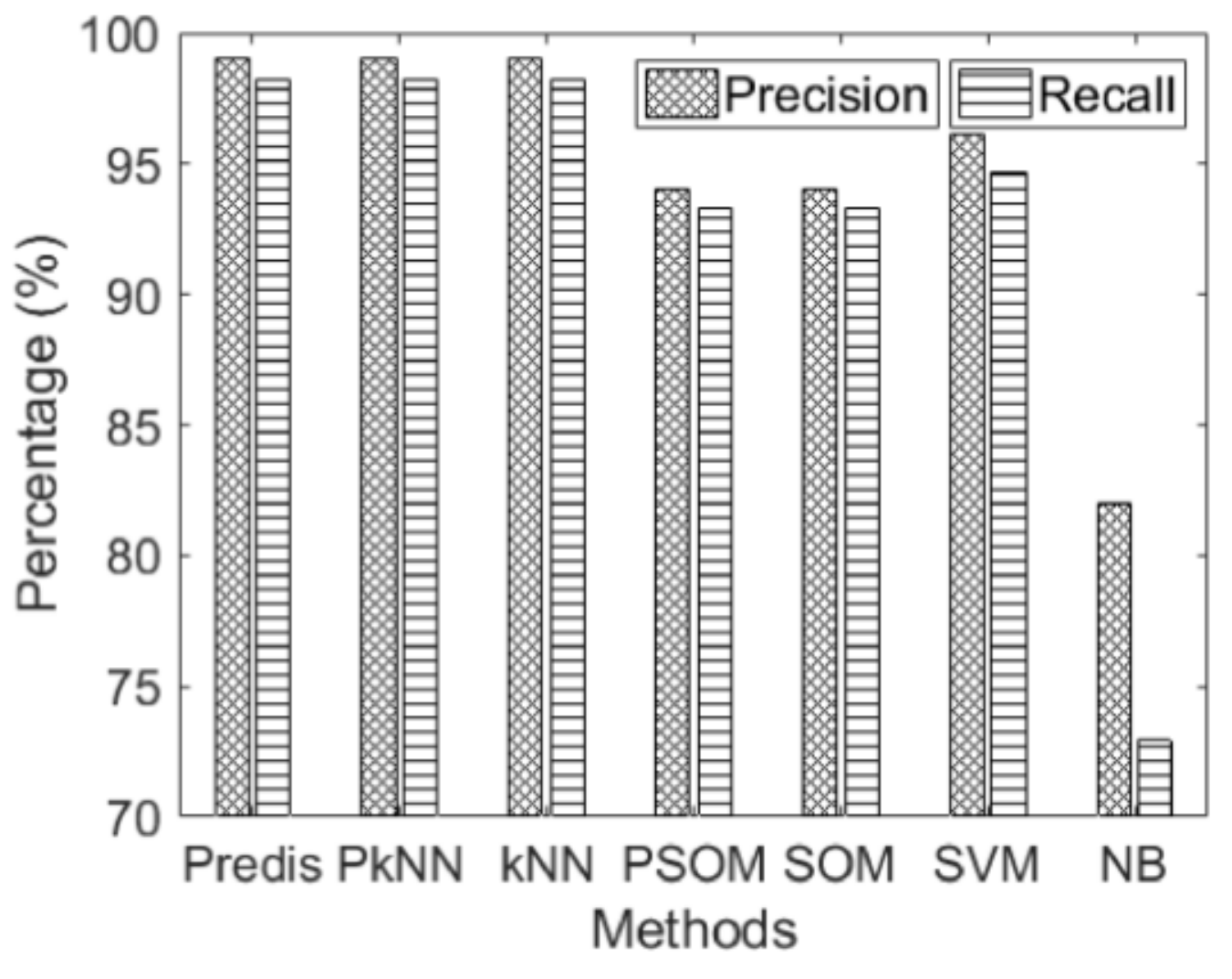}\\
    \scriptsize{(a) Evaluation of Accuracy}
\end{minipage}%
\quad
\begin{minipage}[t]{0.48\textwidth}
    \centering
    \includegraphics[height=5cm]{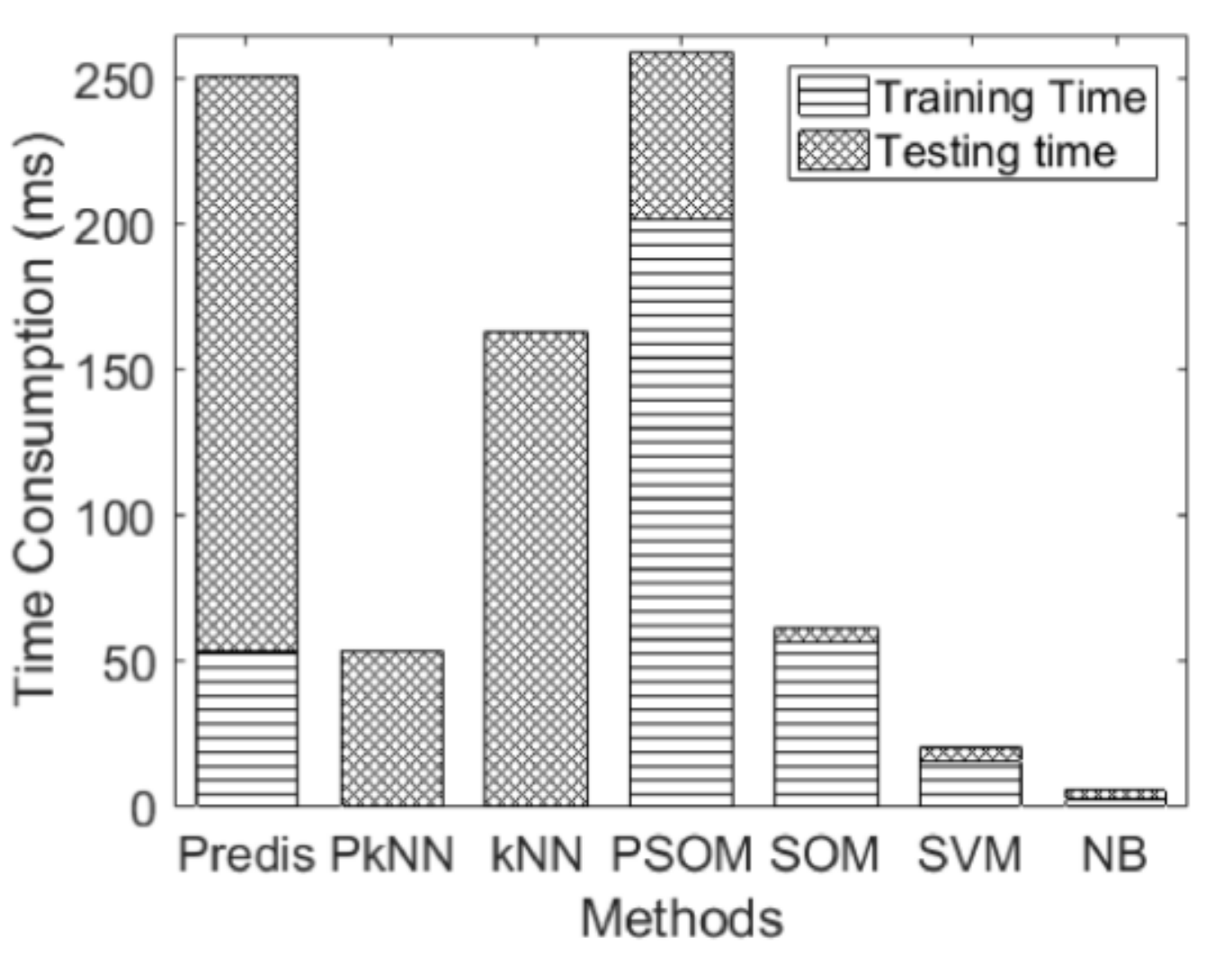}\\
    \scriptsize{(b) Evaluation of Time Consumption}
\end{minipage}%
 \caption{Comprehensive Evaluation of the Classifier in \texttt{Predis}.}
    \label{fig:Comprehensive Evaluation of the Classifier in Predis}
\end{figure*}

\subsection{Evaluation of Accuracy}
\emph{(1) Accuracy evaluation between single-domain and cross-domain scenarios}.

Cross-validation is conducted using Dataset 1.
From the results in Table \ref{table:Accuracy comparison between single domain and cross-domain outlier detection},
we can see that the precision and recall of the single-domain detection is lower than those of the cross-domain detection, which undoubtedly proves our stance that DDoS attacks detection in cross-domain is truly necessary.
We also observe that \texttt{Predis} is superior to PSOM in terms of presicion and recall.
In domain 172.16.115.*, \texttt{Predis} and PSOM both have low precision and recall.
The reason lies in that fewer hosts are invaded by attackers in domain 172.16.115.*, and invaded hosts have not generated many abnormal traffic. The lack of training dataset results in not so outstanding detection results.
But \texttt{Predis}'s precision and recall is still above $0.88$.

\begin{table}[!t]
\centering
\small
\caption{Accuracy Comparison between Single-Domain and Cross- Domain Approaches}\label{table:Accuracy comparison between single domain and cross-domain outlier detection}
\renewcommand{\arraystretch}{0.9}
\begin{tabular}{p{1.7cm}<{\centering}|p{1.2cm}<{\centering}|p{1.2cm}<{\centering}|p{1.2cm}<{\centering}|p{1.2cm}<{\centering}|p{1.2cm}<{\centering}|p{1.2cm}<{\centering}|p{1.2cm}<{\centering}|p{1.2cm}<{\centering}|}
\cline{2-9}
&\multicolumn{4}{c|}{LLDOS 1.0 as Training Dataset}&\multicolumn{4}{c|}{LLDOS 2.0.2 as Training Dataset}\\
\cline{2-9}
&\multicolumn{2}{c|}{Precision}&\multicolumn{2}{c|}{Recall}&\multicolumn{2}{c|}{Precision}&\multicolumn{2}{c|}{Recall}\\
\hline
\multicolumn{1}{|c|}{Domains}&\texttt{Predis}&PSOM&\texttt{Predis}&PSOM&\texttt{Predis}&PSOM&\texttt{Predis}&PSOM\\
\hline
\multicolumn{1}{|c|}{172.16.112.*}&0.9720 &0.9713  &0.9010 &0.5558 	&0.9730 &0.9700 &0.9000 &0.6734 		\\

\multicolumn{1}{|c|}{172.16.113.*}&0.9973 &0.9600   &0.9930 	&0.6700 	&0.9914 &0.9651 &0.9916 &0.8100 \\

\multicolumn{1}{|c|}{172.16.114.*}&0.9900 &0.9843  &0.9874 &0.7800 	&0.9255 &0.7630 &0.8500 &0.7180 		\\

\multicolumn{1}{|c|}{172.16.115.*}&0.8980 &0.8741 &0.8865 &0.8700 &0.8690 &0.8053 &0.8500 &0.8156 \\

\multicolumn{1}{|c|}{Victim}&0.9710 &0.8160 &0.9750 	&0.7071 		&0.8320 &0.8086 &0.7020 &0.6700 	\\	

\multicolumn{1}{|c|}{Attacker}&0.9200 &0.7753  &0.9200 	&0.7200 		&0.9052 &0.4133 &0.8700 &0.6400 	\\	
\hline	
\multicolumn{1}{|c|}{Domains Merge}&0.9985 &0.9780  &0.9923 &0.9653 &0.9920 	&0.8986 &0.9812 &0.8200 		\\
\hline
\end{tabular}
\end{table}

\emph{(2) Accuracy evaluation of detecting DDoS attacks at early stages}.

It is desirable to detect DDoS attacks at the first and second stages.
We conduct cross-validation in Dataset 1 and divided Dataset 1's traces into three stages (scanning, intrusion, and attacking). At both the attack scanning and intrusion stages, \texttt{Predis} delivers excellent detection results as shown in Table \ref{table:Accuracy evaluation of detecting DDoS attacks in its early stage}, which means that \texttt{Predis} can identify attacks at early stages.
In contrast to PSOM, \texttt{Predis} achieves better detection results at any stage of the attack, and its detection result at stages of scanning and intrusion is suboptimal.
Moreover, the precision of \texttt{Predis} is $0.9919$ when LLDOS 1.0 as the training dataset at the attacking stage.

\begin{table}[t]
\centering
\small
\caption{DDoS Attacks Detection Accuracy at Each Stage}\label{table:Accuracy evaluation of detecting DDoS attacks in its early stage}

\renewcommand{\arraystretch}{0.9}
\begin{tabular}{p{1.7cm}<{\centering}|p{1.2cm}<{\centering}|p{1.2cm}<{\centering}|p{1.2cm}<{\centering}|p{1.2cm}<{\centering}|p{1.2cm}<{\centering}|p{1.2cm}<{\centering}|p{1.2cm}<{\centering}|p{1.2cm}<{\centering}|}
\cline{2-9}
&\multicolumn{4}{c|}{LLDOS 1.0 as Training Dataset}&\multicolumn{4}{c|}{LLDOS 2.0.2 as Training Dataset}\\
\cline{2-9}
&\multicolumn{2}{c|}{Precision}&\multicolumn{2}{c|}{Recall}&\multicolumn{2}{c|}{Precision}&\multicolumn{2}{c|}{Recall}\\
\hline
\multicolumn{1}{|c|}{Domains}&\texttt{Predis}&PSOM&\texttt{Predis}&PSOM&\texttt{Predis}&PSOM&\texttt{Predis}&PSOM\\
\hline
\multicolumn{1}{|c|}{Scanning}&0.9734 &0.8730  &0.8645  &0.7630 	&0.8612 &0.8053 &0.9251 &0.8945 		\\
\hline
\multicolumn{1}{|c|}{Intrusion}&0.9920  &0.8740 &0.9900 &0.8821 	&0.9400 &0.8392 &0.8830 &0.8755 	\\
\hline
\multicolumn{1}{|c|}{Attacking}&0.9919  &0.9733  &0.9614 &0.9200 		&0.9808 &0.9472 &0.8792 	&0.8740 		\\
\hline
\end{tabular}
\end{table}

\emph{(3) Accuracy evaluation with privacy-preserving or without.}

Using Dataset 2 for cross-validation, we make a comparison between
\texttt{Predis}, the SVM and SOM methods.
To evaluate the impact of privacy-preserving in \texttt{Predis} on detection accuracy,
we also disable the privacy-preserving function in \texttt{Predis} and refer to this variation as CAD in comparison.

Experimental results are plotted in Figure \ref{fig:Accuracy evaluation with privacy-preserving or without}, where the precision and recall are exhibited in two subfigures,
respectively.
The x-axis indicates the serial number in cross-validation.

When we detail \texttt{Predis} in Section \ref{sec:scheme-attacks detection in DS}, we have proved through theoretical analysis that the detection result of introducing privacy protection will not have any impact on detection accuracy.
This conclusion is validated in Figure \ref{fig:Accuracy evaluation with privacy-preserving or without},
where \texttt{Predis} and CAD obtain the same detection accuracy.
In addition, we can find that \texttt{Predis} outperforms the other two methods in terms of both precision and recall.

\begin{figure*}
\begin{minipage}[t]{0.48\textwidth}
    \centering
    \includegraphics[height=5cm]{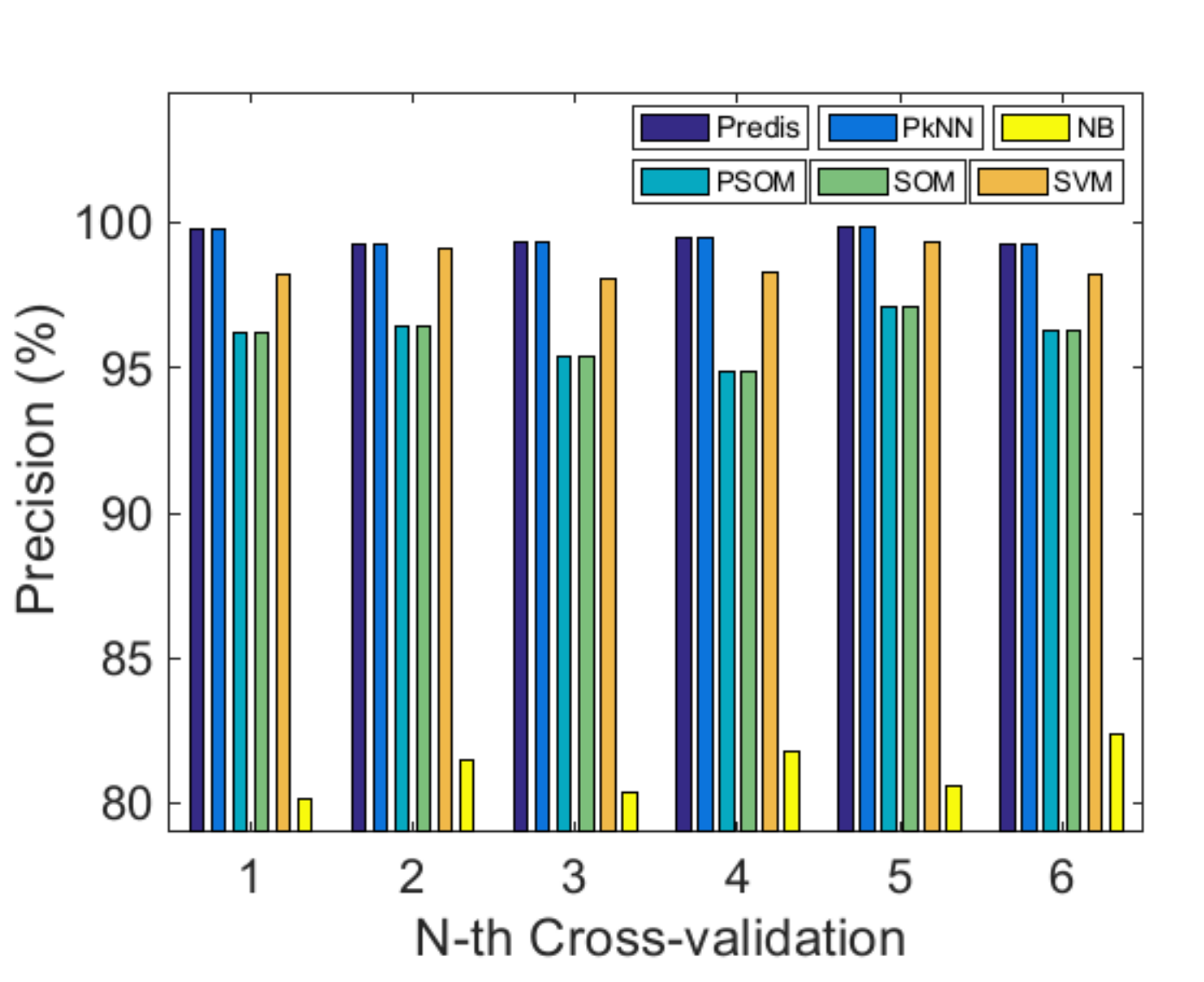}\\
    \scriptsize{(a) Precision}
\end{minipage}%
\quad
\begin{minipage}[t]{0.48\textwidth}
    \centering
    \includegraphics[height=5cm]{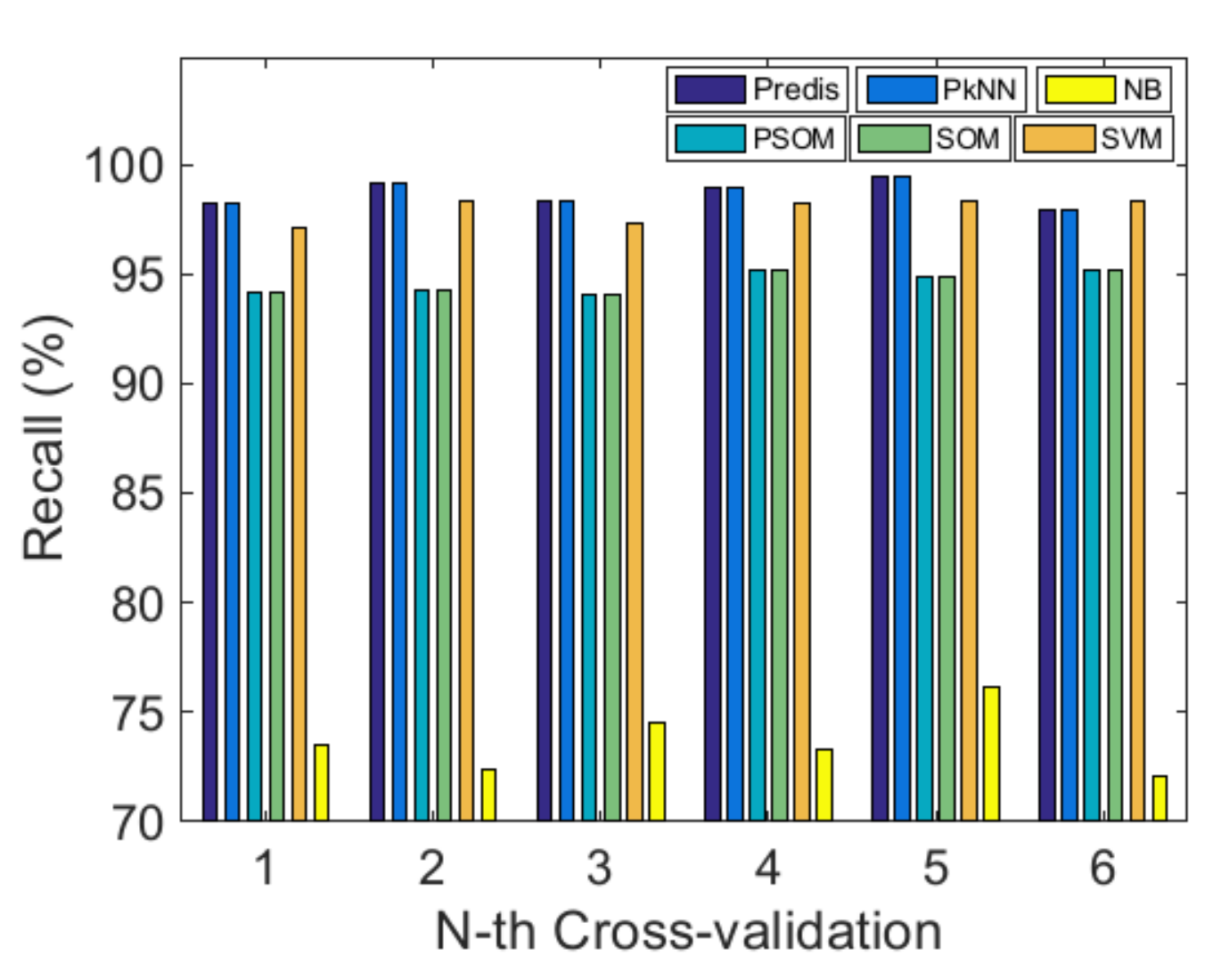}\\
    \scriptsize{(b) Recall}
\end{minipage}%
\caption{The Results of Precision and Recall of $n$-th Cross-Validation Using Dataset 2.}
\label{fig:Accuracy evaluation with privacy-preserving or without}
\end{figure*}

\subsection{Evaluation of Scalability}

Using Dataset 1, we record the changes of time consumption and accuracy when the number of domains goes up from one to six,
as shown in Table \ref{table:Evaluation of expansibility in Dataset 1} and Figure \ref{fig:Evaluation of expansibility}.

When the number of domains increases,
time consumption is only determined by the amount of traffic instances and, the detection effect remains almost unchanged.
The time complexity of the improved kNN is $O(n)$, where $n$ is the number of traning instances.
As can be seen from the experimental results,
the time consumption of \texttt{Predis} does not surge and meets linear variations as the number of domains increases.

\begin{table}[!t]
\centering
\small
\caption{Evaluation of Detection Accuracy with Varied Numbers of Domains Using Dataset 1}\label{table:Evaluation of expansibility in Dataset 1}

\renewcommand{\arraystretch}{0.9}
\begin{tabular}{p{1.7cm}<{\centering}|p{1.2cm}<{\centering}|p{1.2cm}<{\centering}|p{1.2cm}<{\centering}|p{1.2cm}<{\centering}|p{1.2cm}<{\centering}|p{1.2cm}<{\centering}|p{1.2cm}<{\centering}|p{1.2cm}<{\centering}|}
\cline{2-9}
&\multicolumn{4}{c|}{LLDOS 1.0 as Training Dataset}&\multicolumn{4}{c|}{LLDOS 2.0.2 as Training Dataset}\\
\cline{2-9}
&\multicolumn{2}{c|}{Precision}&\multicolumn{2}{c|}{Recall}&\multicolumn{2}{c|}{Precision}&\multicolumn{2}{c|}{Recall}\\
\hline
\multicolumn{1}{|c|}{Number of domains}&\texttt{Predis}&PSOM&\texttt{Predis}&PSOM&\texttt{Predis}&PSOM&\texttt{Predis}&PSOM\\
\hline
\multicolumn{1}{|c|}{1}& 0.9720 &0.9713  & 0.9010 &0.5558 	&0.9730&0.9700&0.9000&0.6734		\\

\multicolumn{1}{|c|}{2}& 0.9840 &0.9672   &0.9769 &0.6530 	&0.9207&0.9270&0.9501&0.9468	\\

\multicolumn{1}{|c|}{3}& 0.9912 & 0.9760 &0.9800 &0.6734 		&0.9232&0.9105&0.9543&0.9208		\\

\multicolumn{1}{|c|}{4}& 0.9914 & 0.9613 &0.9900 &0.7133 &0.9308&0.9256&0.9683&0.9660			\\

\multicolumn{1}{|c|}{5}&0.9920  &0.9514  &0.9913 	&0.8320 		&0.9591&0.8512&0.9537&0.9201	\\	

\multicolumn{1}{|c|}{6}&0.9880 & 0.9438 &0.9864 &0.7689 		&0.9508&0.8380&0.9517&0.9106	\\	
\hline	
\multicolumn{1}{|c|}{Average}& 0.9864 &0.9618 &0.9709 &0.6994 		&0.9424&0.9037&0.9463&0.8896	\\	
\hline
\end{tabular}
\end{table}

\begin{figure*}
\begin{minipage}[t]{0.48\textwidth}
    \centering
    \includegraphics[height=5cm]{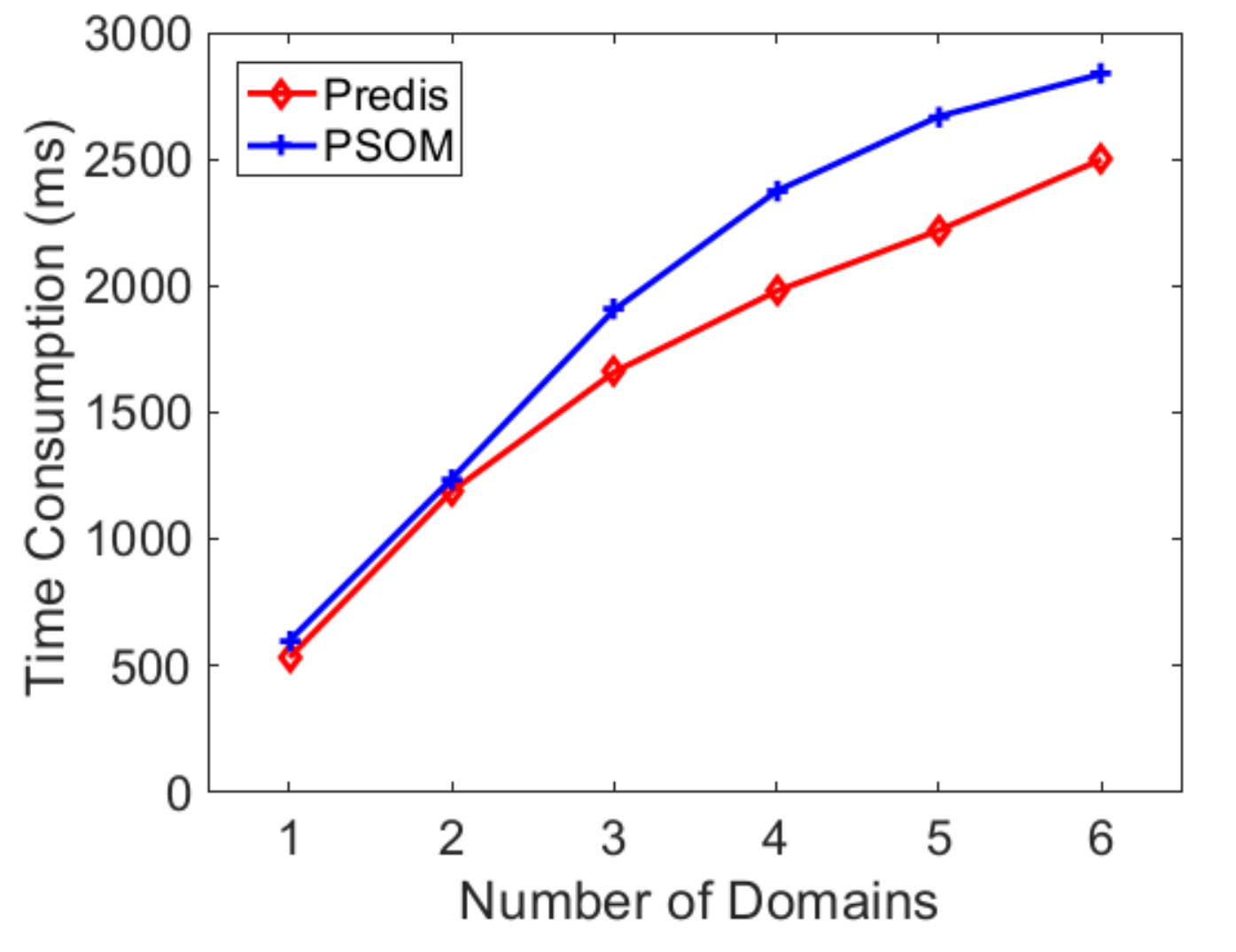}\\
    \scriptsize{(a) LLDOS 1.0 as Training Dataset}
\end{minipage}%
\quad
\begin{minipage}[t]{0.48\textwidth}
    \centering
    \includegraphics[height=5cm]{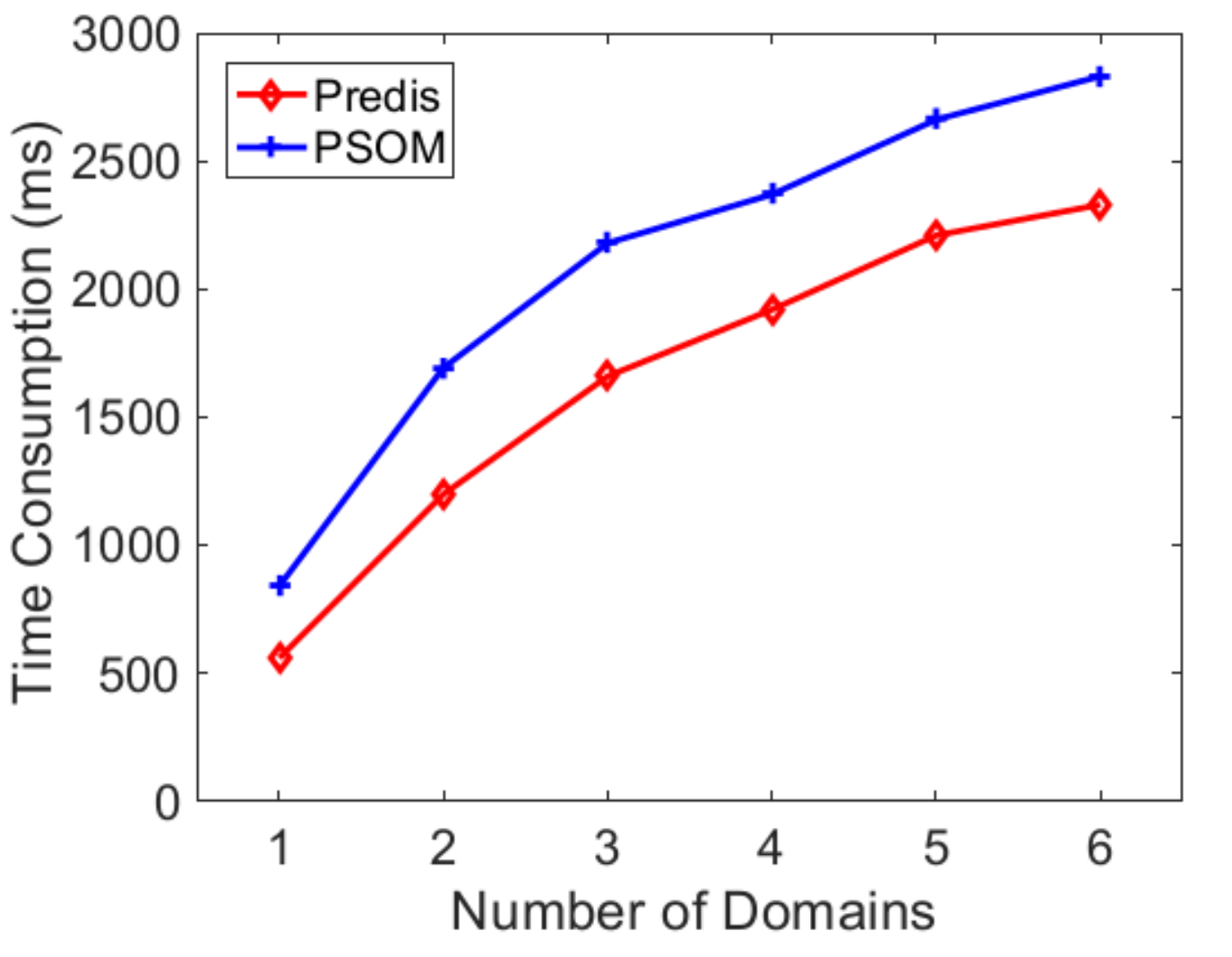}\\
    \scriptsize{(b) LLDOS 2.0.2 as Training Dataset}
\end{minipage}%
\caption{Evaluation of Time Consumption with Varied Numbers of Domains Using Dataset 1.}
    \label{fig:Evaluation of expansibility}
\end{figure*}

\subsection{Evaluation of Time Consumption}
First, from Figure \ref{fig:Evaluation of expansibility}, we can see that with the increasing number of domains, \texttt{Predis} has obvious advantages in terms of time-consumption compared to PSOM.

We analyze the MAWI dataset \cite{48} which are based on the collected network traffic during 7 years of a specific link between Japan and the USA. The backbone generates about 6,000 KB (130,000 flows) traffic per second. If we can process 13,000 flows per second (1/10 of the traffic generated by this backbone line), it shows that \texttt{Predis} can meet the time-consuming requirement.
The time spent is the sum of the time spent between CS and DS.
We record the time spent on two server when the amount of test instances increase from 1,000 to 10,000, where Dataset 2 is used. From Figure \ref{fig:Evaluation of time}(a) we can see that when 10,000 flows for testing, the total time does not exceed 1 second. With privacy-preserving, the time consumption of \texttt{Predis} is also acceptable.

In addition, using Dataset 2, we record the time spent in 6 cross-validation experiments, where PSOM is the control group and the results are shown in Figure \ref{fig:Evaluation of time}(b). We can see in Figure \ref{fig:Evaluation of time}(b), compared to a similar scheme PSOM, \texttt{Predis} has a lower time consumption.

\begin{figure*}
\begin{minipage}[t]{0.48\textwidth}
    \centering
    \includegraphics[height=5cm]{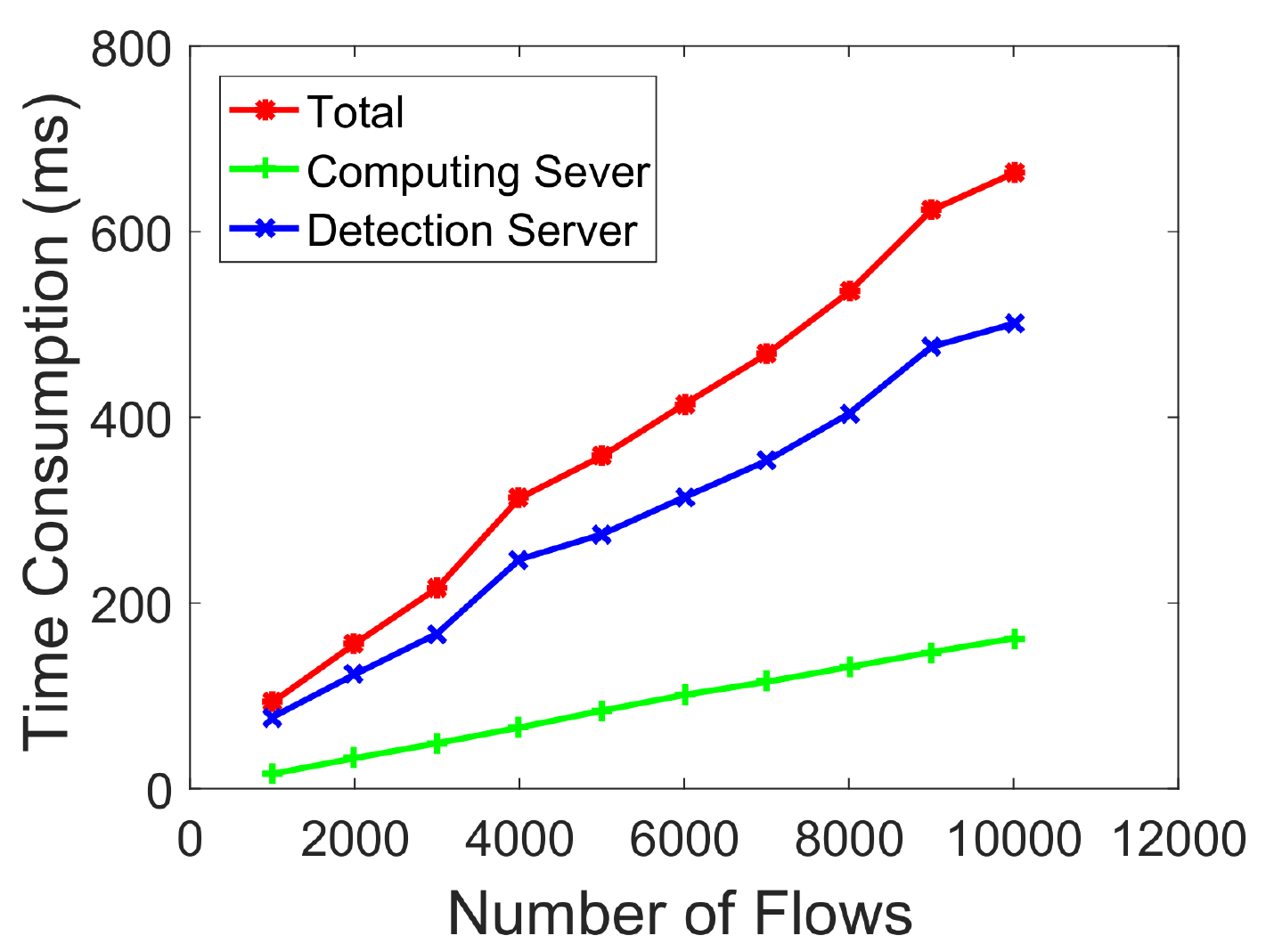}\\
    \scriptsize{(a) Time Consumption with Varied Numbers of Flows}
\end{minipage}%
\quad
\begin{minipage}[t]{0.48\textwidth}
    \centering
    \includegraphics[height=5cm]{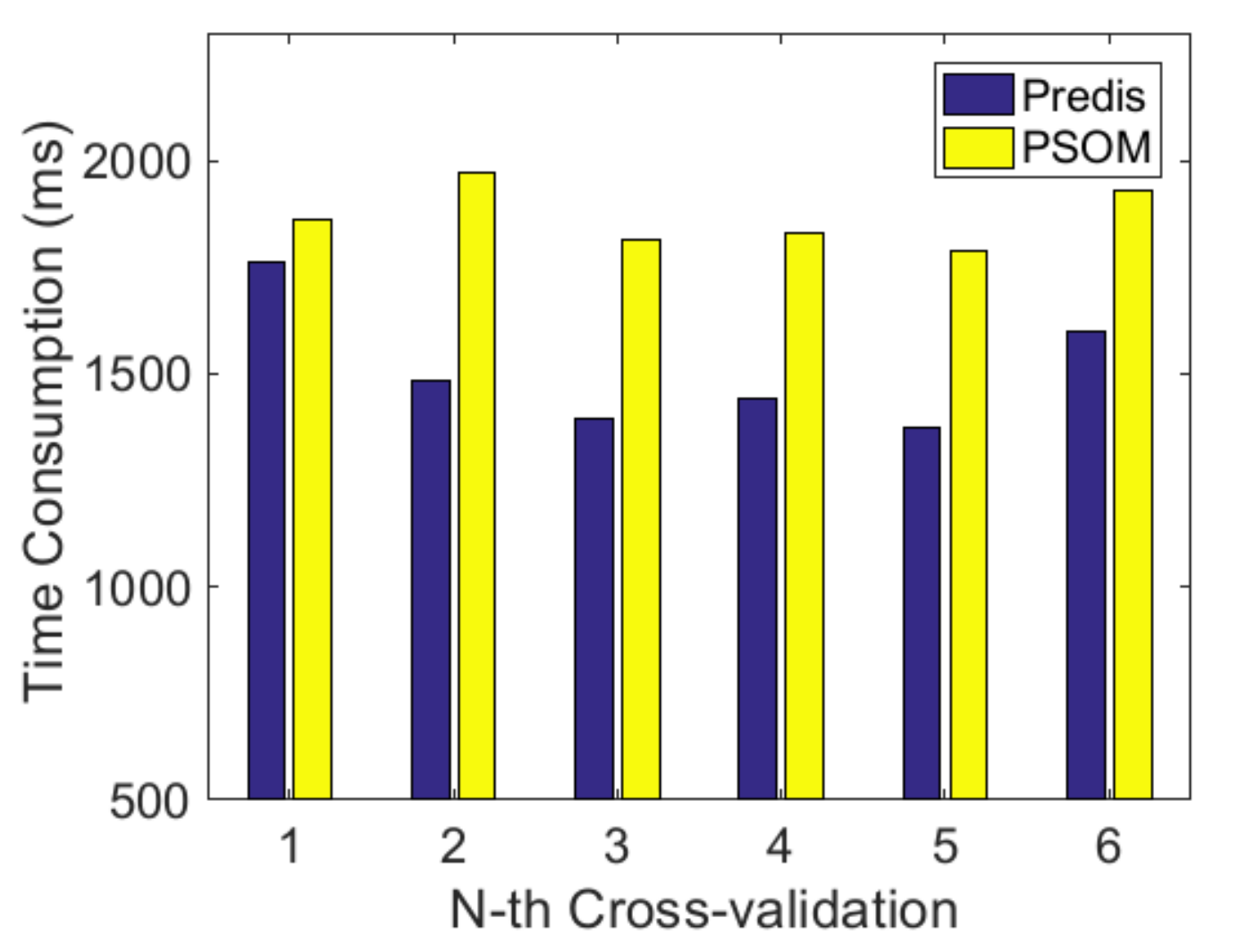}\\
    \scriptsize{(b) Evaluation of Time Consumption Using Dataset 2}
\end{minipage}%
\caption{Evaluation of Time Consumption.}
\label{fig:Evaluation of time}
\end{figure*}

\subsection{Evaluation of Detecting Other Attacks}
We hope that the attacks detection scheme will detect not only DDOS attacks but also other attacks.
We testify through evaluation that \texttt{Predis} is not only suitable for detecting DDOS attacks, but also capable to detect a variety of attacks and retains an excellent accuracy after changing the feature selection model properly.
We conduct experiments using the KDD Cup 1999 traces, which have several other types of attacks besides DDoS attacks.
The feature selection model here is a four-tuple ($Count$, $Src\ bytes$, $Dst\ bytes$, $Logged\ in$).


\emph{(1) Time consumption evaluation when detecting other attacks}.

We detect three types of attacks in the KDD Cup 1999 traces, such as DOS, Prob, and U2R, and record the time it takes as well as accuracy. Results are depicted in Table \ref{table:Time-consuming evaluation when detect other attack}.

\begin{table}[!t]
\centering
\small
\caption{Time Consumption on Detecting Other Attacks}\label{table:Time-consuming evaluation when detect other attack}

\renewcommand{\arraystretch}{0.9}
\begin{tabular}{cc|c|c|c|c|c|}
\cline{2-7}
&\multicolumn{2}{|c|}{DOS}&\multicolumn{2}{c|}{Prob}&\multicolumn{2}{c|}{U2R}\\
\hline
\multicolumn{1}{|c|}{Criterion}&\texttt{Predis}&PSOM&\texttt{Predis}&PSOM&\texttt{Predis}&PSOM\\

\hline
\multicolumn{1}{|c|}{Precision}&0.9161&0.8872&0.9901&0.8753&0.8834&0.8810\\
\hline
\multicolumn{1}{|c|}{Recall}&0.9003&0.8706&0.9874&0.9644&0.7900&0.7650\\
\hline
\multicolumn{1}{|c|}{Total Time}&616ms&707ms&785ms&833ms&569ms&640ms\\
\hline
\end{tabular}
\end{table}

\emph{(2) Accuracy evaluation when detecting other attacks}.

We detect three types of attacks in the KDD Cup 1999 traces, such as DOS, Prob, and U2R, and record the accuracy
in Table \ref{table:Time-Accuracy evaluation when detect other attack},
where SOM and SVM are compared.

\begin{table}[!t]
\centering
\small
\caption{Evaluation of Accuracy in Detecting Other Attacks}\label{table:Time-Accuracy evaluation when detect other attack}

\renewcommand{\arraystretch}{0.9}
\begin{tabular}{cc|c|c|c|c|c|c|c|c|}
\cline{2-10}
&\multicolumn{3}{|c|}{DOS}&\multicolumn{3}{c|}{Prob}&\multicolumn{3}{c|}{U2R}\\
\hline
\multicolumn{1}{|c|}{Criterion}&\texttt{Predis}&SOM&SVM&\texttt{Predis}&SOM&SVM&\texttt{Predis}&SOM&SVM\\
\hline
\multicolumn{1}{|c|}{Precision}&0.9161&0.8872& 0.9024&0.9901&0.8753&0.9732&0.8834&0.8310&0.8604 \\
\hline
\multicolumn{1}{|c|}{Recall}&0.9103&0.8706& 0.9041&0.9874&0.9644&0.9566&0.7900&0.7650&0.7622\\
\hline
\end{tabular}
\end{table}

When detecting attacks of DoS, Prob, and U2R, the precision of \texttt{Predis} is no less than 90\%.
Compared to the same privacy-preserving scheme PSOM, \texttt{Predis} is better in terms of time consumption (as shown in Table \ref{table:Time-consuming evaluation when detect other attack}).
Because in the KDD Cup 1999 traces, U2R attacks have very few abnormal traffic instances, therefore recall is relative low when testing U2R attacks.
Compared with SVM and SOM, \texttt{Predis} wins out as far as precision and recall are concerned (as shown in Table ref{table:Time-Accuracy evaluation when detect other attack}). Through these two experiments, it has been proven that \texttt{Predis} not only can detect DDoS attacks excellently, but also can detect other attacks very well when appropriately selected features are incorporated.

\section{Discussion}\label{sec:Discussion}
If SDNs controllers can know the DDoS attacks in it early stages in time and make corresponding measures (e.g., limiting the SYN / ICMP traffic, filtering specific IP addresses, traffic cleaning, etc.), SDNs controllers can prevent the DDoS attacks before it causes damage.
SDNs uses centralized control for traffic forwarding mechanism that makes it easier to stop DDoS attacks.
\texttt{Predis} has been demonstrated that it can detect DDoS attacks in the early stages of an attack (as show in Table \ref{table:Accuracy evaluation of detecting DDoS attacks in its early stage}).
Once the server detects a DDoS attacks, it alerts the controller in time.
The controller immediately responds to the alert, and prevents any further damage from the attack.

The roles in \texttt{Predis} include two servers and SDNs controllers.
\texttt{Predis} can prevent the collusion between one or more domains with one server, so the two servers can be respectively deployed in any one of the participating domains, or wherever providing secure communications with the domains over the TLS protocol.
The data plane and the control plane of the SDNs communicate by using a control-data-plane interface (CDPI).
The main use of the uniform communication standard is OpenFlow protocol.
Flow table operations that include flow table pretreatment, communicate with the server, proper handling for DDoS alerts or other operation can be deployed and implemented through the use of the southbound interfaces (i.e., COPI).

Due to the development of different network access technologies and different communication systems, resource scheduling and fusion in heterogeneous networks have become hot topics.
SDNs can achieve unified management in configuration heterogeneous network equipment and it open a variety of interfaces.
But these interfaces perhaps is exploited by attackers, such as tapping, interception and DDoS attacks.
We can regard the heterogeneous networks as a domain under the unified management of a SDNs.
Using \texttt{Predis}, Users can achieve secure cross-domain DDoS attacks detection and resist the threats from DDoS attacks in heterogeneous networks.
In addition, \texttt{Predis}'s idea of combining data perturbation with data encryption to provide cross-domain privacy protection may also be transplanted to other application scenarios in the future to achieve other secure multi-party computing.
Overlay networks such as P2P (Peer-to-Peer) add virtual channels to physical networks to enhance network flexibility.
Each node in P2P networks may be the data provider.
When considering monitoring the traffic of multiple nodes without compromising the privacy, each node in the P2P can be considered as a domain, and the idea of \texttt{Predis} in cross-domain detection can be used to provide security monitoring service.

\section{Conclusion}\label{sec:conclusion}
In this paper, we presented a SDN-based cross-domain attacks detection scheme with privacy protection.
We studied cross-domain privacy protection problems and DDoS attacks detection based on SDNs.
We combined geometric transformation and data encryption method with the view to protect privacies.
We broke down the detection process into two steps, disturbance and detection, and introduced two servers that work together to complete the detection process.
We optimized the kNN for low time consumption and high accuracy. Extensively experiment results showed that \texttt{Predis} is capable to detect cross-domain anomalies while preserving the privacy with low time consumption and high accuracy.
We plan to further reduce the time consumption of \texttt{Predis} in attacks detection in the future work.

\label{sec:reference}
\bibliographystyle{bare_jrnl}
\bibliography{bare_jrnl}

\vspace{-60pt}
\begin{IEEEbiographynophoto}
{Liehuang Zhu} is a professor in the Department of Computer Science at Beijing Institute of Technology.
He is selected into the Program for New Century Excellent Talents in University from Ministry of Education, P.R. China.
His research interests include Internet of Things, Cloud Computing Security, Internet and Mobile Security.
\end{IEEEbiographynophoto}

\vspace{-60pt}

\begin{IEEEbiographynophoto}
{Xiangyun Tang} received the B.Eng degree in computer science from Minzu University of China, Beijing, China in 2016.
Currently she is a M.S. student in the Department of Computer Science, Beijing Institute of Technology.
Her research interest is Secure in Software Defined Networks.
\end{IEEEbiographynophoto}

\vspace{-60pt}

\begin{IEEEbiographynophoto}
{Meng Shen} received the B.Eng degree from Shandong University, Jinan, China in 2009, and the Ph.D degree from Tsinghua University, Beijing, China in 2014, both in computer science. Currently he serves in Beijing Institute of Technology, Beijing, China, as an assistant professor. His research interests include privacy protection of cloud-based services, network virtualization and traffic engineering.
He received the Best Paper Runner-Up Award at IEEE IPCCC 2014.
He is a member of the IEEE.
\end{IEEEbiographynophoto}

\vspace{-60pt}

\begin{IEEEbiographynophoto}
{Xiaojiang Du} is a tenured professor in the Department of Computer and Information Sciences at Temple University, Philadelphia, USA.
His research interests are wireless communications, wireless networks, security, and systems.
He has authored over 230 journal and conference papers in these areas, as well as a book published by Springer.
Dr. Du has been awarded more than \$5 million US dollars research grants from the US National Science Foundation (NSF), Army Research Office, Air Force Research Lab, NASA, the State of Pennsylvania, and Amazon.
He serves on the editorial boards of three international journals.
Dr. Du is a Senior Member of IEEE and a Life Member of ACM.
\end{IEEEbiographynophoto}

\vspace{-60pt}

\begin{IEEEbiographynophoto}
{Mohsen Guizani} has received his B.S., M.S., and Ph.D. from Syracuse University.
He is currently a professor and the Electrical and Computer Engineering Department Chair at the University of Idaho.
His research interests include wireless communications,
mobile cloud computing, computer networks, security, and smart grid.
He is the author of nine books and more than 400 publications. He was the Chair of the IEEE Communications Society Wireless Technical Committee.
He served as an IEEE Computer Society Distinguished Speaker.
\end{IEEEbiographynophoto}

\end{document}